\documentclass[twocolumn]{aastex63}
\shorttitle{Stratified streaming instabilities}
\shortauthors{M.-K.\ Lin}

\usepackage{amsmath}
\usepackage{paralist}
\usepackage{url}
\usepackage{bm}
\usepackage{multirow}
\usepackage[T1]{fontenc}
\usepackage[utf8]{inputenc}
\newcommand{\p}{\partial}
\newcommand{\zmax}{z_\mathrm{max}}

\newcommand{\ii}{\mathrm{i}}
\newcommand{\C}{\mathcal{C}^\lambda}

\newcommand{\dln}[1]{(\ln{#1})^\prime}

\newcommand{\dd}{\delta}
\newcommand{\real}[1]{\operatorname{Re}\left(#1\right)}
\newcommand{\imag}[1]{\operatorname{Im}\left(#1\right)}

\newcommand{\Hdust}{H_\mathrm{d}}
\newcommand{\Hgas}{H_\mathrm{g}}
\newcommand{\hgas}{h_\mathrm{g}}

\newcommand{\rhod}{\rho_\mathrm{d}}
\newcommand{\rhog}{\rho_\mathrm{g}}
\newcommand{\Vd}{\bm{V}_\mathrm{d}}
\newcommand{\Vg}{\bm{V}_\mathrm{g}}
\newcommand{\vd}{\bm{v}_\mathrm{d}}
\newcommand{\vg}{\bm{v}_\mathrm{g}}
\newcommand{\vdx}{v_{\mathrm{d}x}}
\newcommand{\vdy}{v_{\mathrm{d}y}}
\newcommand{\vdz}{v_{\mathrm{d}z}}
\newcommand{\vgx}{v_{\mathrm{g}x}}
\newcommand{\vgy}{v_{\mathrm{g}y}}
\newcommand{\vgz}{v_{\mathrm{g}z}}
\newcommand{\sigd}{\Sigma_\mathrm{d}}
\newcommand{\sigg}{\Sigma_\mathrm{g}}
\newcommand{\tstop}{t_\mathrm{s}}

\newcommand{\taus}{\tau_\mathrm{s}}
\newcommand{\st}{\mathrm{St}}
\newcommand{\w}{\widetilde{W}}
\newcommand{\fdust}{f_\mathrm{d}}
\newcommand{\fg}{f_\mathrm{g}}

\newcommand{\tcool}{t_\mathrm{cool}}
\newcommand{\delrhod}{\frac{\dd\rhod}{\rhod}}
\newcommand{\delrhog}{\frac{\dd\rhog}{\rhog}}
\newcommand{\delrho}{\frac{\dd\rho}{\rho}}
\newcommand{\delK}{\frac{\dd \mathcal{K}}{\mathcal{K}}}
\newcommand{\calK}{\mathcal{K}}
\newcommand{\deleps}{\dd\epsilon}
\newcommand{\bmv}{\bm{v}}
\newcommand{\re}{\operatorname{Re}}
\newcommand{\im}{\operatorname{Im}}
\defcitealias{lin12}{L12}
\defcitealias{lin15}{LY15}
\defcitealias{nelson13}{N13}
\defcitealias{barker15}{BL15}
\defcitealias{mcnally14}{MP14}
\defcitealias{goldreich67}{GS67}
\defcitealias{youdin05a}{YG05}

\begin{document}
\title{{Stratified and vertically-shearing streaming instabilities in protoplanetary disks}}
\author{Min-Kai Lin}
\affiliation{Institute of Astronomy and Astrophysics, Academia Sinica, Taipei 10617, Taiwan}
\email{mklin@asiaa.sinica.edu.tw}

\begin{abstract}
Under the right conditions, the streaming instability between {imperfectly coupled} dust and gas is a powerful mechanism for planetesimal formation as it can concentrate dust grains to the point of gravitational collapse. In its simplest form, the {streaming} instability can be captured {by analyzing the linear stability of}
unstratified disk models{, which} represent the midplane of protoplanetary disks. {We extend such studies by} carrying out vertically-global linear stability analyses of dust layers in protoplanetary disks.
We find the dominant {form of} instability in stratified dust layers is {one} driven by the vertical gradient in the rotation velocity of the dust-gas mixture{, but also requires partial dust-gas coupling. These vertically-shearing streaming} instabilities grow on orbital timescales and
occur on radial length scales $\sim10^{-3}\Hgas$, where $\Hgas$ is the local pressure scale height.
The classic streaming instability, {associated with} the relative radial drift between dust and gas, occur on radial length scales $\sim10^{-2}\Hgas$, but have much smaller growth rates than {vertically-shearing streaming} instabilities. Including gas viscosity is strongly stabilizing and leads to vertically-elongated disturbances. We briefly discuss {the potential effects of vertically-shearing streaming instabilities on} planetesimal formation.  
\end{abstract}

\section{Introduction}





The formation of planetesimals from mm--cm sized dust grains or pebbles in protoplanetary disks (PPDs) is a key stage in planet formation {\citep{birnstiel16}}. Neither pair-wise collisions nor gravitational forces lead to effective growth on pebble scales \citep{chiang10,blum18}. However, if a swarm of solids can be made sufficiently dense relative to the ambient gas, then it can undergo direct self-gravitational collapse into km or larger-sized planetesimals  \citep{goldreich73}. The critical dust-to-gas ratio for collapse {is $\gg 1$ \citep{shi13}. This should be} compared to the typical value of $\sim 1\%$ expected {uniformly throughout a} newly-born PPD \citep{testi14}. Thus, an efficient mechanism is needed to first concentrate dust grains.
{These include dust settling, particle, trapping by pressure bumps, and dust-gas instabilities \citep{johansen14}.}

The streaming instability \citep[SI,][]{youdin05a,youdin07} is one such candidate. The SI is linear instability in rotating flows of partially-coupled dust and gas that mutually interact through frictional drag -- conditions natural in PPDs --- which can amplify dust-to-gas ratios by orders of magnitude and trigger gravitational collapse \citep{johansen09}, although the SI itself does not require self-gravity. {In the often considered, idealized case of a laminar disk with a monodisperse dust population, the} SI is a robust process and has thus received considerable attention as the de facto mechanism for planetesimal formation.


The SI has undergone intense studies through numerical simulations \citep{johansen07,bai10b,yang14,yang17}. Modern simulations have generalized the SI to consider magnetic fields  \citep{balsara09,tilley11,yang18}; various geometries \citep{kowalik13,schreiber18}; multiple grain sizes  \citep{bai10,bai10c,schaffer18,llambay19,krapp19,zhu20}; turbulence \citep{schafer20,gole20}; self-gravity \citep{simon16,schafer17,li19}, pressure bumps \citep{carrera20}, etc. These efforts are necessary to understand  planetesimal formation in realistic PPDs. Indeed, sophisticated simulations show that planetesimals formed through the SI have properties consistent with that in the solar system \citep{nesvorny19}.


On the other hand, a physical understanding of the SI through analytical studies has progressed more slowly. \cite{jacquet11} showed that the SI is driven by a process of runaway dust-trapping by pressure bumps, while \cite{lin17} gave a thermodynamic interpretation of the SI in which {partial dust-gas coupling leads to} `PdV' work {that} acts to amplify oscillations.

{Recently, \cite{squire20} presented detailed models of \cite{youdin05a}'s classic SI, in which the mutual interaction between epicyclic motions and the relative radial dust-gas drift in a PPD leads to growing perturbations. 
In fact, for small dust-to-gas ratios, the classic SI belong to}
a broader class of `resonant drag instabilities' \citep{squire18a,squire18b,zhuravlev19} generic to dusty-gas in which {the relative dust-gas motions resonate with a wave mode in the gas. The classic SI is thus expected in PPDs since dust and gas naturally exhibit a relative radial drift, as} the gas rotation is partially supported by a (negative) radial pressure gradient \citep{whipple72,weiden77}.

The classic SI can be captured in relatively simple disk models, such as that employed by \citeauthor{youdin05a}, who considered a small region near the disk midplane. In this limit, the vertical gravity from the central star can be neglected, which produces a uniform vertical disk structure. These `unstratified' disk models allow significant simplifications for analyses of the linear SI {and generalizations thereof, which  include the effect of pressure bumps, multiple species, and turbulence}
\citep{jaupart20,auffinger18,krapp19,paardekooper20,chen20,umurhan20,pan20b}.


However, PPDs do have {a} vertical structure. Dust-settling, which leads to the formation of a dense particle layer about the disk midplane
 \citep{dubruelle95}, is often considered as a prerequisite to trigger the SI, as it requires dust-to-gas ratios {of order unity or above} to operate efficiently \citep{youdin05a}. Dust-settling naturally produces a stratified dust layer.  In fact, stratified simulations
are now common,
but the linear SI has not been examined in stratified disks. Filling this gap will be helpful in understanding how the SI operates in realistic PPDs.

The purpose of this work is to generalize previous studies of the linear SI to account for the vertical structure of dust layers in PPDs. To this end, we analyze the stability of vertically-global, radially-local models of dusty disks. We employ the standard, two-fluid description of dusty-gas, as well as a simplified `one-fluid' model \citep{lin17} to verify some of our calculations.

Our main result is that in stratified disks the vertical variation in the azimuthal velocity of the dusty-gas (or vertical shear) provides a significant source of {`free energy' that can be accessed by partial dust-gas coupling, which results in instability.
We typically find these vertically-shearing streaming instabilities (VSSIs)} dominate over classic SIs, so the former should be the first to develop in settled dust layers. Our study confirms and expands upon an earlier work from \cite{ishitsu09}, who used direct simulations to study the effect of vertical dust density gradients on the evolution of dust layers.

This paper is organized as follows. By way of introducing notation, we first provide  order-of-magnitude motivations to examine the SI in stratified disks in \S\ref{motivation}. We then describe our framework and disk models in \S\ref{basic}. Results from our linear stability analyses are presented in \S\ref{results} for three examples:
{a high dust density layer, a low dust density layer}, and a viscous disk. We discuss implications of our findings in \S\ref{discussion} and conclude in \S\ref{summary}. {A list of frequently used symbols is summarized in Appendix \ref{appendix_symbols}.}


\section{Physical Motivation}\label{motivation}

\subsection{Geometric considerations}

The classic SI of \cite{youdin05a}, discovered in unstratified disk models, is driven by the relative radial drift between dust and gas,
\begin{align}\label{vdrift}
    v_\mathrm{drift} = - \frac{2\st (1+\epsilon) \eta r \Omega}{\st^2 + (1 + \epsilon)^2}
\end{align}
\citep{nakagawa86}, where the Stokes number $\st$
is a dimensionless inverse measure of the strength of dust-gas coupling (and is proportional to the grain size), $\epsilon$ is the dust-to-gas volume density ratio, $r$ is the cylindrical distance from the star, $\Omega = \sqrt{GM_*/r^3}$ is the Keplerian frequency ($M_*$ and $G$ being the stellar mass and gravitational constant, respectively), and $\eta$ is a dimensionless measure of the global radial pressure gradient defined as
\begin{align}\label{eta_def}
  \eta \equiv - \frac{1}{2r\Omega^2\rhog}\frac{\p P}{\p r},
\end{align}
where $\rhog$ is the gas density and $P$ is the gas pressure. PPDs have $\eta \sim O(\hgas^2)$, where $\hgas\equiv \Hgas/r$ is the gas disk aspect-ratio and $\Hgas$ is the pressure scale-height, with $\hgas\sim 0.05$.

We thus expect the SI to have characteristic lengthscales $\sim \eta r$, which is significantly shorter than the global radial lengthscales of typical PPD disk models ($\sim r$). The SI can thus be considered as a radially-localized phenomenon. However, the situation differs in the vertical direction.

In realistic PPDs, dust settles into a layer of thickness $\Hdust$, which can be related to the midplane dust-to-gas mass density ratio $\epsilon_0$ and the metallicity
\begin{align} \label{metal_def}
    Z \equiv \frac{\Sigma_\mathrm{d}}{\Sigma_\mathrm{g}}\simeq  \epsilon_0 \frac{\Hdust}{\Hgas}
\end{align}
\citep{johansen14}, where $\Sigma_\text{d,g}$ are the dust and gas surface densities, respectively. The SI operates on dynamical timescales when $\epsilon_0 \gtrsim 1$ \citep{youdin05a}. To meet this condition at standard solar solid abundances of $Z\simeq 0.01$, dust layers should be thin, $\Hdust\simeq 0.01\Hgas$. For a SI mode with vertical lengthscale $\eta r $ to exist, it should fit inside the dust layer, or
\begin{align}
    \chi\equiv \frac{\eta r}{\Hdust} = \hat{\eta}\left(\frac{\epsilon_0}{Z}\right)\lesssim 1,  
\end{align}
where $\hat{\eta} \equiv \eta/\hgas$ and PPDs typically have $\hat{\eta}\simeq 0.05$.

{For} a settled dust layer with $\epsilon_0\simeq 1$ in a disk with standard  metallicity $Z\simeq 0.01$ we find $\chi \simeq 5${, violating the above condition.}
{Moreover, when gas viscosity and particle diffusion are considered, SI modes have vertical lengthscales comparable to $\Hgas$ \citep{umurhan20}; while further restricting vertical lengthscales to $\lesssim \Hdust$ results in negligible growth rates \citep[][]{chen20}. These findings call for stratified analyses.}


\subsection{Energetic considerations}

Dusty PPDs possess vertical shear: the settled, dust-rich midplane rotates closer to the Keplerian speed $r\Omega$ than the gas-dominated, pressure-supported disk away from the dust layer, which has a sub-Keplerian rotation of $ (1-\eta)r\Omega$ because $\eta > 0$ {usually}. Such a vertical variation of the disk's rotation speed is an important source of free energy (as borne out of our calculations).

{Now, the difference in the azimuthal velocity of the dusty midplane and the overlaying gas is $\Delta v_\phi\sim \eta r\Omega$. For a dust layer thickness $\Hdust$ we can thus estimate the vertical shear rate within the layer as $\eta r \Omega/\Hdust = \chi \Omega$.} For small grains, this vertical shear rate is larger than the relative radial drift rate $v_\mathrm{drift}/\eta r$ by a factor of $\st^{-1}\gg 1$.

{We can also compare vertical shear to the vertical settling of grains, as the latter can trigger a `dust settling instability' \citep[DSI,][]{squire18b,krapp20}. Taking the typical settling speed of a dust grain to be  $\left|v_\mathrm{dz}\right|\sim \st \Hdust\Omega$ \citep{takeuchi02}, we find $\left|\Delta  v_\phi/ v_\mathrm{dz}\right|\sim \hat{\eta}\epsilon_0/(\st Z)$. For PPDs, with $\epsilon_0\sim 1$, $\hat{\eta}\sim Z \sim O(10^{-2})$, this ratio is $\sim \st^{-1}$, i.e. large for small grains.
}

{The above estimates suggest that for small grains,  vertical shear is a much larger energy source than the  relative radial drift or dust settling. Indeed, in the limit of $\st \to 0$ the classic SI and DSI are suppressed and vertical shear drives non-axisymmetric Kelvin-Helmholtz instabilities \citep[KHI,][]{chiang08,lee10}.}

In this work we consider axisymmetric disturbances so that KHIs are not applicable. \cite{lin17} showed that axisymmetric dusty disks are generally stable in the limit $\st\to 0$, however large the vertical shear rate. This is due to stabilization by dust-induced, effective buoyancy forces \citep[see also][]{lin19}.

However, the above result ceases to be valid for $\st \neq 0$, because in {this} case dust and gas are no longer perfectly coupled {and they can stream past one another}, which diminishes the stabilizing effect of dusty buoyancy. This is similar to how rapid cooling can enable the `vertical shear instability' {(VSI)} in gaseous PPDs \citep{nelson13,lin15} {by eliminating gas buoyancy. (For the gaseous VSI, vertical shear originates from the disk's radial thermal structure.)}

We {can} therefore expect in a stratified, dusty disk the free energy associated with vertical shear{, here a result of dust settling},
{to be accessible} through an instability with {non-vanishing} particle sizes.
{Indeed, we will find such instabilities
are the dominant modes in stratified disks. We refer to them as vertically-shearing streaming instabilities (VSSIs), since both vertical shear and dust-gas streaming motions are necessary.}



\section{Basic equations}\label{basic}

We consider a non-self-gravitating, unmagnetized PPD comprised of gas and a single species of dust grains around a central star of mass $M_*$. The gas component has density, pressure, and velocity fields  $(\rhog, P, \Vg)$. We assume an isothermal gas so that $P=c_s^2\rhog$ with a constant sound-speed $c_s=\Hgas\Omega$.

We treat the dust population as a pressureless fluid with density and velocity $(\rhod,\Vd)$. The dust and gas fluids interact via a drag force parameterized by a stopping time $\taus$ (see below). The fluid approximation for dust is then valid for well-coupled, small grains  such that  $\taus\lesssim\Omega^{-1}$  \citep{jacquet11}. 


\subsection{Two-fluid, radially local, axisymmetric disk model}

We study radially-localized disturbances in the aforementioned dusty disk using the
shearing box framework \citep{goldreich65}. The shearing box is centered about
a fiducial point $(r_0, \phi_0, 0)$ in cylindrical co-ordinates on the star, which rotates at the reference Keplerian frequency  $\Omega(r_0)\equiv\Omega_0$, i.e. $\phi_0 = \Omega_0 t$. Cartesian co-ordinates $(x,y,z)$ in the box correspond to the radial, azimuthal, and vertical directions in the global disk. The radial extent of the box is assumed to be much smaller than $r_0$, so that curvature terms from the cylindrical geometry can be neglected. Keplerian rotation {is then approximated} as the linear shear flow $-(3/2)\Omega_0 x \hat{\bm{y}}$. We define $\bm{v}_\mathrm{d,g} \equiv \bm{V}_\mathrm{d,g} - (r_0 - 3x/2)\Omega_0\hat{\bm{y}}$ as the local dust and gas velocities in the shearing box relative to this linear shear flow. We assume axisymmetry throughout, so that $\p_y \equiv 0$.

The governing equations for the dust component in the shearing box are
\begin{align}
    &  \frac{\p\rhod}{\p t} + \nabla\cdot\left(\rhod\vd\right) =  \nabla\cdot\left[D\rhog\nabla\left(\frac{\rhod}{\rhog}\right)\right],\label{dust_mass}\\
    &  \frac{\p\vd}{\p t} + \vd\cdot\nabla\vd = 2\Omega_0\vdy \hat{\bm{x}} -
  \frac{\Omega_0}{2} \vdx \hat{\bm{y}} - \Omega_0^2 z\hat{\bm{z}}\notag\\
  & \phantom{\frac{\p\vd}{\p t} + \vd\cdot\nabla\vd =}
   -\frac{1}{\taus}(\vd-\vg), \label{dust_mom}
\end{align}
where $D$ is a constant diffusion coefficient {defined below}. The third term on the right-hand-side (RHS) of Eq. \ref{dust_mom} corresponds to the vertical component of the stellar gravity  in the thin-disk limit. The last term on the RHS corresponds to gas drag, the strength of which is characterized by {the} stopping time $\taus$.

For the gas, we include the effect of a global radial pressure gradient in the shearing box by writing
\begin{align}
    \nabla P \to \nabla P - 2\eta_0 r_0 \Omega_0^2\rhog \hat{\bm{x}},
\end{align}
where $\eta_0 = \eta(r=r_0,z=0)$ and  $\eta$ is defined by Eq. \ref{eta_def}. That is, the global radial pressure gradient is modeled as a constant forcing. We then re-interpret $P$ as pressure fluctuations in the shearing box. The governing equations for the gas component are then
\begin{align}
&  \frac{\p\rhog}{\p t} + \nabla\cdot\left(\rhog\vg\right)= 0,\label{gas_mass}\\
&  \frac{\p\vg}{\p t} + \vg\cdot\nabla\vg = 2\Omega_0 \vgy \hat{\bm{x}} -
  \frac{\Omega_0}{2} \vgx \hat{\bm{y}} - \frac{\nabla  P}{\rhog}  \notag\\
&  \phantom{
    \frac{\p\vg}{\p t} + \vg\cdot\nabla\vg =
}
  + 2\eta_0 \Omega_0^2 r_0 \hat{\bm{x}} + \frac{1}{\rhog}\nabla\cdot\bm{T} \notag\\
&  \phantom{
   \frac{\p\vg}{\p t} + \vg\cdot\nabla\vg =
}
 -\Omega_0^2 z \hat{\bm{z}} - \frac{\epsilon}{\taus}(\vg - \vd). \label{gas_mom}
\end{align}
The fifth term on the RHS of Eq. \ref{gas_mom} represent viscous forces, where
\begin{align}\label{stress_tensor}
    \bm{T} = \rhog\nu \left[\nabla\vg+\left(\nabla\vg\right)^\dagger - \frac{2}{3}\bm{I}\nabla\cdot\vg\right]
\end{align}
is the viscous stress tensor and $\nu$ is a kinematic viscosity, prescribed later. The final term on the RHS is the back-reaction of dust drag onto the gas.

The basic equations {\ref{dust_mass}--\ref{dust_mom} and \ref{gas_mass}--\ref{gas_mom}} extend those used by \cite{chen20} with the addition of vertical gravity{, which themselves are  extensions of that in \cite{youdin07b} with the addition of dust diffusion and gas viscosity. We solve {Eqs.  \ref{dust_mass}--\ref{dust_mom} and \ref{gas_mass}--\ref{gas_mom}}  in full to obtain equilibrium states, then solve their linearized versions to study the stability of said equilibria. Both problems are one-dimensional in $z$. For numerical solutions we consider the half-disk $z\in[0,z_\mathrm{max}]$ by imposing symmetry conditions at the midplane. Details are given in  \S\ref{vert_eqm}--\ref{horiz_eqm} and \S\ref{linear}.} Hereafter we drop the subscript `0' for clarity. Below, $\Hgas$ refers to the pressure scale height at the reference radius.

\subsection{Dust-gas drag}
The stopping time $\taus$ is the timescale for a dust particle to reach velocity equilibrium with its surrounding gas. In this work we take $\taus$ to be a constant parameter for simplicity.
It is convenient to define the dimensionless stopping time or Stokes number,
\begin{align}
\st = \taus \Omega.
\end{align}
We consider well-coupled, or small dust grains with $\st \ll 1$.

Physically, {$\st$} depends on the particle and gas properties, such as grain size ($a$), internal density ($\rho_\bullet$), and gas density \citep{weiden77}. To put our calculations in context, consider grains in the Epstein regime in a Minimum Mass Solar Nebula-like disk (MMSN) described in \cite{chiang10}. We then find

\begin{align}
    \st = 0.019 F^{-1}\left(\frac{r}{30\text{au}}\right)^{3/2}\left(\frac{\rho_\bullet}{\text{g}\text{cm}^{-3}}\right)\left(\frac{a}{\mathrm{mm}}\right),
\end{align}
where $F$ is a mass scale relative to the standard MMSN ($F=1$). We are mostly interested in mm or sub-mm-sized grains with internal density $1\text{g}\text{cm}^{-3}$ at tens of au in MMSN-like disks.

\subsection{Dust diffusion}

We include dust diffusion to allow a stratified equilibrium state to be defined, in which dust settling is balanced by dust diffusion. Without dust diffusion, particles would continuously settle and no steady state can be established for standard stability analyses. Dust diffusion is usually attributed to gas turbulence \citep[e.g.][]{youdin07,laibe20}{, which is often modeled as a gas viscosity. We thus parameterize dust diffusion in terms of a gas viscosity,  although for the most part we neglect viscosity in the gas equations.}

We model dust diffusion via the constant parameter $\delta$ such that
\begin{align}
  D = \delta c_s \Hgas, \label{diffusion_coefficient}
\end{align}
with $\delta$ given by
\begin{align}
\delta = \frac{1 + \st + 4\st^2}{\left(1+\st^2\right)^2}\alpha\label{delta_alpha}
\end{align}
\citep{youdin07,youdin11}, where $\alpha$ is an input constant turbulent viscosity parameter defined below. In practice, $\delta \simeq \alpha$ since we consider small grains.

\subsection{Turbulent viscosity}
When considered, we model gas turbulence via a viscous stress tensor (see Eq. \ref{stress_tensor}) and adopt the standard alpha prescription \citep{shakura73} such that the kinematic viscosity is
\begin{align}
    \nu = \alpha c_s \Hgas \frac{\rho_\mathrm{g,eqm}}{\rhog},
\end{align}
where $\rho_\mathrm{g,eqm}(z)$ denotes the equilibrium gas density, derived below. We use this prescription so that the dynamic viscosity $\rhog\nu$ is a fixed function of space, which avoids viscous overstabilities that could complicate results \citep{latter06,lin16}.

\subsection{Two-fluid equilibria}\label{2feqm}
We seek steady, horizontally uniform equilibria with $\p_t=\p_x = 0$. The gas continuity equation then imply $\vgz=0$. All other are quantities are non-zero and $z$-dependent, e.g. $\rhog = \rho_\mathrm{g,eqm}(z)$. For clarity, hereafter we drop the subscript `eqm' on the equilibrium fields.

\subsubsection{Vertical equilibrium}\label{vert_eqm}

The equilibrium dust mass and vertical momentum equations are:
\begin{align}
    \frac{d\ln{\epsilon}}{dz} &= \frac{\vdz}{D},\\
    c_s^2 \frac{d\ln{\rhog}}{dz} &= \frac{\epsilon\Omega}{\st}\vdz - \Omega^2 z,\\
    \vdz\frac{d\vdz}{dz} &= - \Omega^2 z - \frac{\Omega}{\st}\vdz,
\end{align}
{where we recall $\epsilon = \rhod/\rhog$.}
For constant {$\st$} these may be solved exactly to yield
\begin{align}
    &\epsilon(z) = \epsilon_0 \exp{\left( - \frac{\beta}{2\delta}\frac{z^2}{\Hgas^2}\right)},\label{2f_eps}\\
    &\rhog(z) = \rho_{\mathrm{g}0}\exp{\left[ \frac{\delta}{\st}\left(\epsilon - \epsilon_0\right) -
    \frac{z^2}{2\Hgas^2}\right]},\label{2f_rhog}\\
    & \vdz(z) = -\beta z\Omega\label{2f_vdz},
\end{align}
where $\epsilon_0$, $\rho_{\mathrm{g}0}$ is the mid-plane dust-to-gas ratio and gas density, respectively; and
\begin{align}\label{beta_def}
    \beta \equiv \frac{1}{2\st}\left(1 - \sqrt{1 - 4\st^2}\right).
\end{align}
We thus require $\st\leq 0.5$. Note that $\beta\simeq \st$\footnote{{This can been seen by performing a Taylor expasion of the numerator in Eq. \ref{beta_def}}} for the small particles considered in this work ($\st\ll 1$). We also consider weak diffusion such that $\delta\ll \st$.
The dust layer thickness can then be approximated by
\begin{align}\label{Hdust_def}
    \Hdust  = \sqrt{\frac{\delta}{\delta + \st}}\Hgas
\end{align}
\citep{dubruelle95,zhu15}. {
The local metallicity is
\begin{align}
    Z \equiv \frac{\int_{-\infty}^\infty\rho_\mathrm{d}dz}{\int_{-\infty}^\infty\rho_\mathrm{g}dz} = \frac{\sigd}{\sigg}.
\end{align}
In practice, we adjust $\epsilon_0$ until a specified value of $Z$ is obtained. However, Eq. \ref{metal_def} and Eq. \ref{Hdust_def} also gives an adequate estimate,  $\epsilon_0\simeq Z\sqrt{\st/\delta}$. The vertical structure is then completely determined.
}


\subsubsection{Horizontal equilibrium}\label{horiz_eqm}

The equilibrium horizontal momentum equations are
\begin{align}
    & \vdz \frac{d\vdx}{dz} = 2\Omega\vdy - \frac{\Omega}{\st}\left(\vdx - \vgx\right),\\
    & \vdz\frac{d\vdy}{dz} = -\frac{\Omega}{2}\vdx - \frac{\Omega}{\st}\left(\vdy - \vgy\right), \\
    & 0 = 2\Omega \vgy + 2\eta\Omega^2 r  - \frac{\epsilon\Omega}{\st}\left(\vgx - \vdx \right)
    + \frac{\nu}{\rhog}\frac{d}{dz}\left(\rhog \frac{d\vgx}{dz}\right),\\
    & 0 = -\frac{\Omega}{2} \vgx - \frac{\epsilon\Omega}{\st}\left(\vgy - \vdy \right)
    + \frac{\nu}{\rhog}\frac{d}{dz}\left(\rhog \frac{d\vgy}{dz}\right),
\end{align}
{with $\epsilon(z)$, $\rhog(z)$, and $\vdz(z)$ given by Eqs. \ref{2f_eps} -- \ref{2f_vdz}.} The horizontal velocity profiles must, in general, be solved numerically subject to appropriate boundary conditions. However, for $|z|\to\infty$ and thus $\epsilon\to 0$, the
dust and gas equations decouple and we obtain
\begin{align}
   &\lim_{\epsilon\to 0}\vdx = -\frac{2\st\eta r\Omega}{1+\st^2},\label{dust_eqm_bc1}\\
    &\lim_{\epsilon\to 0}\vdy = -\frac{\eta r\Omega}{1+\st^2},\label{dust_eqm_bc2} \\
&\lim_{\epsilon\to 0} \vgx = 0,\label{gas_eqm_bc1}\\
&\lim_{\epsilon\to 0} \vgy = - \eta r \Omega,\label{gas_eqm_bc2}
\end{align}
which are constants. These correspond to a sub-Keplerian gas flow that does not feel the dust drag, while the dust drifts inwards in response to gas drag. Eqs. \ref{dust_eqm_bc1}--\ref{gas_eqm_bc2} are consistent with the unstratified solutions of \cite{nakagawa86}. 

When gas viscosity is neglected, we impose Eqs. \ref{dust_eqm_bc1}--\ref{dust_eqm_bc2} at a finite height $z=\zmax$ {such that} $\epsilon \ll 1$. When gas viscosity is included, we impose Eqs. \ref{dust_eqm_bc1}--\ref{gas_eqm_bc2} at $z=\zmax$, as well as $\vgx^\prime(0) = \vgy^\prime(0) = 0$, where $^\prime$ denotes $d/dz$.

\subsection{One-fluid models}

We also employ the `one-fluid' description of dusty gas \citep{laibe14,price15,lin17} to confirm selected results. In this framework, we work with the total mass $\rho$ and the center-of-mass velocity $\bm{v}_c$ of the dust-plus-gas mixture, which is treated as a single, ideal fluid subject to a special cooling function. This approximation is valid for small particles with $\st\ll 1$. Our one-fluid formulation includes dust diffusion, but without gas viscosity \citep[cf.][]{lovascio19}. Details are given in Appendix \ref{one_fluid_model}.

\section{Linear problem}\label{linear}


We consider axisymmetric Eulerian perturbations of the form
\begin{align}\label{euler_pert}
  \delta\rhog(z)\exp{\left( \ii k_xx +
    \sigma t\right)},
\end{align}
where $k_x$ is a (real) radial wavenumber taken to be positive without loss of generality; and $\sigma$ is the complex frequency {or eigenvalue},
\begin{align}
\sigma \equiv s - \ii\omega,
\end{align}
where $s$ is the real growth rate and $\omega$ is the oscillation frequency. {We also refer to the complex amplitudes such as $\delta\rhog(z)$ and their normalized versions (e.g. $\delta\rhog/\rhog$) as the eigenfunctions. The initial perturbation in real space is then obtained by taking $\re\left[\delta\rhog\exp{\left(\ii k_x x\right)}\right]$. Similar definitions apply to other variables.}

The linearized equations for the dust fluid read:
\begin{align}
&\sigma \frac{\dd\rhod}{\rhod} + \ii k_x \left(\vdx\delrhod + \dd\vdx\right)
+ \frac{\rhod^\prime}{\rhod}\left(\vdz\delrhod + \dd\vdz\right)\notag \\
& \vdz\left(\delrhod\right)^\prime + \vdz^\prime\delrhod + \dd\vdz^\prime
= -Dk_x^2\frac{\dd\epsilon}{\epsilon} \notag\\
& + \frac{D}{\epsilon}\left[ \frac{\rhog^\prime}{\rhog}\left(\epsilon^\prime\delrhog +
\dd\epsilon^\prime\right)
+\epsilon^{\prime\prime}\delrhog + \epsilon^\prime\left(\delrhog\right)^\prime
+ \dd\epsilon^{\prime\prime}\right],\\
&\sigma \dd\vdx + \ii k_x \vdx\dd\vdx + \vdx^\prime \dd\vdz  + \vdz\dd\vdx^\prime\notag\\
&= 2\Omega\dd\vdy - \frac{\Omega}{\st}\left(\dd\vdx - \dd\vgx\right),\\
&\sigma \dd\vdy + \ii k_x \vdx\dd\vdy  + \vdy^\prime \dd\vdz + \vdz\dd\vdy^\prime\notag\\
&= - \frac{\Omega}{2}\dd\vdx - \frac{\Omega}{\st}\left(\dd\vdy - \dd\vgy\right),\\
&\sigma \dd\vdz + \ii k_x \vdx\dd\vdz  + \vdz^\prime \dd\vdz + \vdz\dd\vdz^\prime\notag\\
&= - \frac{\Omega}{\st}\left(\dd\vdz - \dd\vgz\right),
  \end{align}
and that for the gas equations are:
  \begin{align}
   &\sigma\delrhog + \ii k_x \left(\vgx\delrhog + \dd\vgx\right)
   + \frac{\rhog^\prime}{\rhog}\dd\vgz + \dd\vgz^\prime = 0,\label{lin_gas_mass}\\
 &\sigma\dd\vgx + \ii k_x \vgx\dd\vgx + \vgx^\prime \dd\vgz \notag \\
 &=2\Omega\dd\vgy - \ii k_xc_s^2\delrhog
 + \dd F^\mathrm{br}_x + \delta F^\text{visc}_x,\\
 &\sigma\dd\vgy + \ii k_x \vgx\dd\vgy + \vgy^\prime \dd\vgz
 =-\frac{\Omega}{2}\dd\vgx + \dd F^\mathrm{br}_y + \delta F^\text{visc}_y, \\
& \sigma\dd\vgz + \ii k_x \vgx\dd\vgz
 =-c_s^2\left(\delrhog\right)^\prime + \dd F^\mathrm{br}_z + \delta F^\text{visc}_z,
\end{align}
where the linearized back-reaction force is
\begin{align}
\dd\bm{F}^\mathrm{br} \equiv -\frac{\epsilon\Omega}{\st}\left[\left(\vg - \vd\right)\frac{\deleps}{\epsilon} + \left(\dd\vg - \dd\vd\right)\right],
\end{align}
and the components of the linearized viscous forces are
\begin{align}
\delta F^\text{visc}_x =& \nu \left[\dd\vgx^{\prime\prime} -\frac{4}{3}k_x^2\dd\vgx + \frac{1}{3}\ii k_x\dd\vgz^\prime \right.\notag\\
&\phantom{\nu \left[\right]}\left.+ \frac{\rhog^\prime}{\rhog} \left(\dd\vgx^\prime + \ii k_x \dd\vgz\right)\right] \notag\\
&-\nu \left(\vgx^{\prime\prime} + \frac{\rhog^\prime}{\rhog}\vgx^\prime\right)\delrhog,\label{dFvisc_x}\\
\delta F^\text{visc}_y =& \nu \left[\dd\vgy^{\prime\prime}
+\frac{\rhog^\prime}{\rhog}\dd\vgy^\prime-k_x^2\dd\vgy \right]\notag\\
&-\nu \left(\vgy^{\prime\prime} + \frac{\rhog^\prime}{\rhog}\vgy^\prime\right)\delrhog,\label{dFvisc_y}\\
\delta F^\text{visc}_z = & \nu \left[\frac{4}{3}\dd\vgz^{\prime\prime} - k_x^2\dd\vgz + \frac{1}{3}\ii k_x\dd\vgx^\prime \right.\notag\\
&\phantom{\nu \left[\right]}
\left.+ \frac{\rhog^\prime}{\rhog}\left(\frac{4}{3}\dd\vgz^\prime - \frac{2}{3}\ii k_x \dd\vgx\right)\right]\label{dFvisc_z}
\end{align}
\citep{lin16}. We remark that one can differentiate the gas continuity equation (\ref{lin_gas_mass}) to
eliminate $\dd\vgz^{\prime\prime}$ from the expression of $\dd F^\mathrm{visc}_z$.

In practice, we solve for the perturbation to the dust-to-gas ratio instead of the dust density perturbation, which are related by
\begin{align}
 \delrhod = \frac{\deleps}{\epsilon} + \delrhog \equiv Q + W.
\end{align}
Note that for the strictly isothermal gas we consider $W \equiv \delta\rhog/\rhog = \dd P/P$.

{The linearized equations may be written in the form
\begin{align}\label{eigen_problem}
    \mathcal{L}\bm{g} = \sigma\bm{g},
\end{align}
where $\mathcal{L}$ is an $8\times8$ matrix of linear differential operators and $\bm{g}=\left[W, \dd\vg, Q, \dd\vd \right]^\dagger$. When supplemented with appropriate boundary conditions (see below) this constitutes an eigenvalue problem for $\mathcal{L}$.
}

\subsection{Boundary conditions}
We consider modes symmetric about the mid-plane such that
\begin{align}
    W^\prime(0) = Q^\prime(0) = \dd v_{\mathrm{c},x}^\prime(0)=\dd v_{\mathrm{c},y}^\prime(0) = \dd v_{\mathrm{c},z} (0) =0,
\end{align}
where $\bm{v}_\mathrm{c}$
is the center of mass velocity, see Eq. \ref{com_1fluid}. At the top boundary we impose
\begin{align}
    W^\prime(\zmax) = Q(\zmax) = 0.
\end{align}
When gas viscosity is included we additionally impose
\begin{align}
    \dd \vgx^\prime(0) = \dd \vgy^\prime(0) = \dd \vgx^\prime(\zmax) = \dd \vgy^\prime(\zmax) = 0.
\end{align}

We {generally} find dominant modes have {amplitudes that maximize off the disk-midplane and decay towards the domain boundaries}, as found by \cite{ishitsu09} in direct simulations.
As such, boundary conditions are unlikely to {modify our main  findings.}

\subsection{Numerical method}
We use \textsc{dedalus} \citep{burns19}, a general-purpose spectral code for solving partial differential equations, {including} the linear eigenvalue problem described above. The {eigenfunctions}  are expanded in Chebyshev polynomials $T_n$ up to order $n=N-1$, and the domain is discretized into $N$ points corresponding to the roots of $T_{N}$. Unless otherwise stated, we {take $z_\mathrm{max}\simeq 5\Hdust$}.

We use $N=1024$ for computing the background disk structure\footnote{In the one-fluid formulation we use approximate analytic equilibrium solutions, see Appendix \ref{one_fluid_model}.} and $N=384$ for the linearized equations. {For the latter, \textsc{dedalus} transforms Eq. \ref{eigen_problem} into a generalized matrix eigenvalue problem and solves it via the SciPy package \citep[see Section 9D of][]{burns19}. This directly yields the eigenvalues $\sigma$ and the associated eigenfunctions.}

{For the eigenvalue problem we also use the  \textsc{eigentools}\footnote{\url{https://bitbucket.org/jsoishi/eigentools}.} package to filter out spurious numerical solutions due to the discretization. This is done by comparing eigenvalues obtained  from different vertical resolutions and only keeping those
within some tolerance (here $10^{-6}$), i.e. only physical solutions that converge with respect to $N_z$ are kept. See \cite{barker15} for a similar treatment. We also filter out unphysical solutions with large growth rates compared to $\Omega$ \citep{lin15}.}


{Example source codes for used this work may be obtained from the author's GitHub repository\footnote{\url{https://github.com/minkailin/stratsi}.}.
}

\subsection{Units and notation}

We use normalized units such that $c_s=\Hgas = \Omega = 1$, and quote the dimensionless radial wavenumber $K_x \equiv k_x\Hgas$. It turns out that only the reduced pressure-gradient parameter $\hat{\eta} = \eta r/\Hgas$ is relevant. Eigenfunctions are normalized such that  $\delta\rhod/\rhod = 1$ at its maximum amplitude. For clarity, in plot labels we drop the subscripts `d', `g', and `c' when collectively referring to the dust, gas, and center-of-mass velocity fields.

{We quote $\delta$ to distinguish models in which only dust  diffusion is included, from models wherein corresponding viscous terms are also included in the gas momentum equations, in which case we quote $\alpha$.}

\section{Results}\label{results}

{We present results for a high dust density layer (\S\ref{caseA}), a low dust density layer (\S\ref{caseB}), and a viscous disk (\S\ref{caseC}). For a given set of disk parameters and $K_x$, solving the linearized equations accounts for all vertical structures permitted by the boundary conditions and the finite resolution. This can result in a large number of modes. We are interested in unstable modes as they will dominate over decaying ones in a real disk. Thus, solutions with $s < 0$ are discarded and we focus on those with the largest growth rate $s=s_\mathrm{max}$ at a given $K_x$. Table \ref{vssi_modes} lists, for each case,  approximately\footnote{Due to the finite range and sampling in $K_x$-space.} the most unstable mode over $10^2\leq K_x\leq 10^4$ (based on two-fluid calculations).
}


\begin{deluxetable*}{lcccccclll}
\tablecaption{{Selected unstable modes in stratified dusty disks.} \label{vssi_modes}}
\tablehead{\colhead{Case} & \colhead{$\hat{\eta}$} & \colhead{$Z$}& \colhead{$\st$} & \colhead{$\alpha (\simeq\delta)$} & \colhead{Viscosity\tablenotemark{a}} & \colhead{$k_x\Hgas/10^{3}$} & \colhead{$s_\mathrm{max}/\Omega$} & \colhead{$\omega/\Omega$} & \colhead{Comment}}
\startdata
\multirow{3}{*}{A} & 0.05 & 0.03 & $10^{-2}$ &  $10^{-6}$ &  no &  $3.593814$ &  $1.155043$ & $-0.8250615$ &  high dust density layer, Fig. \ref{caseA_growth_max} \\
& 0.01 & 0.03 & $10^{-2}$ &  $10^{-6}$ &  no &  $5.994843$ & $0.3267454$  &  $-0.1839234$  &  smaller pressure gradient, Fig. \ref{caseA_compare_eta}\\
&0.1 & 0.03 & $10^{-2}$ & $10^{-6}$  &  no &  $3.593814$ & $1.724942$  &  $-1.858679$ &  larger pressure gradient, Fig. \ref{caseA_compare_eta}\\
&0.05 & 0.03 & $10^{-3}$ & $10^{-6}$  &  no &  $10$ &  $0.3901075$ &  $-0.05611060$ &  smaller particle, Fig. \ref{caseA_compare_stokes}\\
& 0.05 & 0.03 & $10^{-1}$ & $10^{-6}$  &  no &  $0.8254042$ & $1.463588$  & $-4.087298$  & larger particle, Fig. \ref{caseA_compare_stokes}\\
& 0.05 & 0.01 & $10^{-2}$ & $10^{-6}$  &  no &  $5.994843$ & $0.9145120$ & $-0.4717328$ & smaller metallicity, Fig. \ref{caseA_compare_Z}\\
& 0.05 & 0.1 & $10^{-2}$ & $10^{-6}$  &  no &  $3.593814$ & $1.308362$  &  $-1.703580$ & larger metallicity, Fig. \ref{caseA_compare_Z}\\
\hline
B & 0.05 & 0.03 & $10^{-3}$ &  $10^{-5}$ &  no &  $10$ &  $0.06512705$ & $-0.01026995$ & low dust density layer, Fig. \ref{caseB_growth_max}\\
\hline
C & 0.05 & 0.01 & $10^{-2}$ &  $10^{-7}$ &  yes &  $1.112355$ &  $0.6007165$ &  $-0.9712429$ & viscous disk, Fig. \ref{caseC_growth}
\enddata
\tablenotetext{a}{Denotes whether or not viscous terms are included the gas momentum equations.}
\end{deluxetable*}

\subsection{Case A: {High dust density layer}}\label{caseA}
\begin{figure}
    \includegraphics[width=\linewidth]{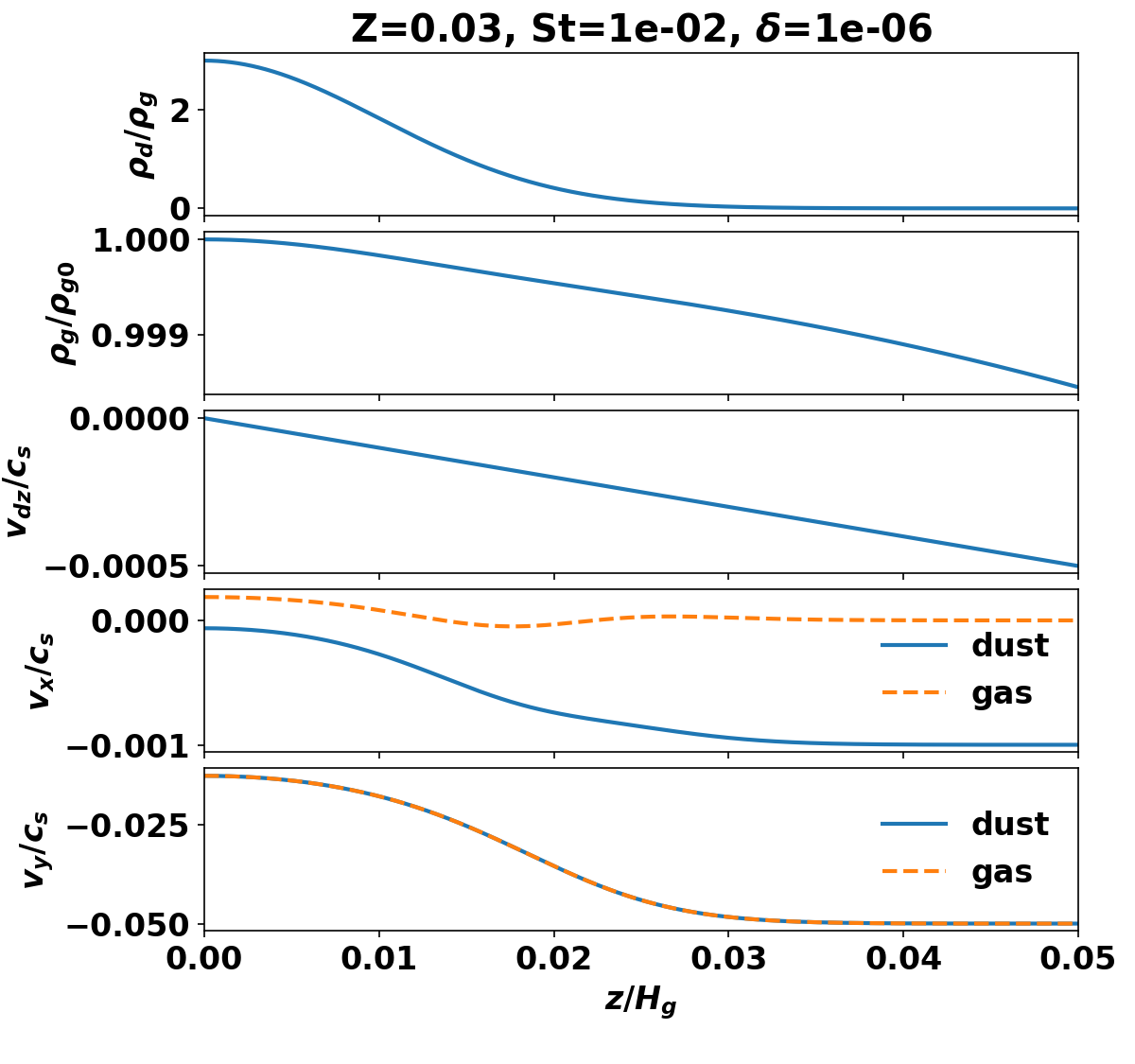}
    \caption{Two-fluid equilibrium for case A {with a high dust density layer}. From top to bottom: dust-to-gas ratio, gas density, vertical dust velocity, radial velocities, and azimuthal velocities.
    \label{caseA_eqm}}
\end{figure}

We first present a fiducial case with well-settled dust. Here, we neglect gas viscosity but include particle diffusion. This enables a comparison between the two-fluid and one-fluid frameworks, since the latter does not include gas viscosity. We choose $Z=0.03$, $\hat{\eta}=0.05$, $\st=10^{-2}$, and $\delta \simeq 10^{-6}$. This gives $\Hdust\simeq 0.01\Hgas$. The equilibrium disk profiles are shown in Fig. \ref{caseA_eqm}.


\begin{figure}
    \includegraphics[width=\linewidth]{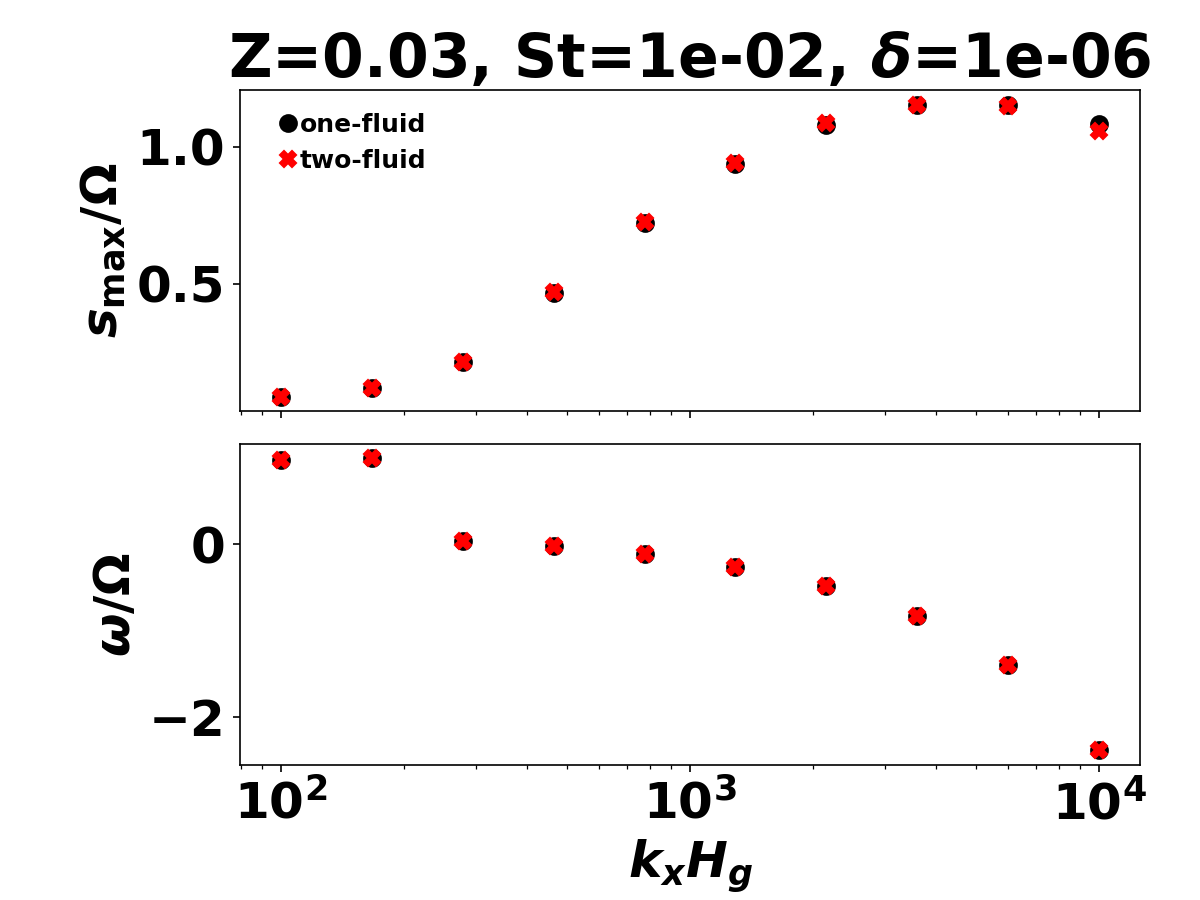}
    \caption{Maximum growth rate (top) and corresponding oscillation frequency (bottom) for unstable modes in case A {(high dust density layer)}, as a function of the dimensionless radial wavenumber $K_x$.
    \label{caseA_growth_max}}
\end{figure}

Fig. \ref{caseA_growth_max} shows the maximum growth rate and corresponding frequencies for case A for $K_x = 100$ to $10^4$. Growth rates increase with $K_x$ and maximizes around $K_x\sim 3600$. We find two classes of unstable modes: for $K_x\lesssim 200$, the oscillation frequency $\omega\sim \Omega$ and is constant; while for $K_x\gtrsim 200$ oscillation frequencies are negative and increase in magnitude. We find good agreement between the one- and two-fluid results{, giving confidence that these are physical solutions.}

\begin{figure*}
    \includegraphics[width=.5\linewidth]{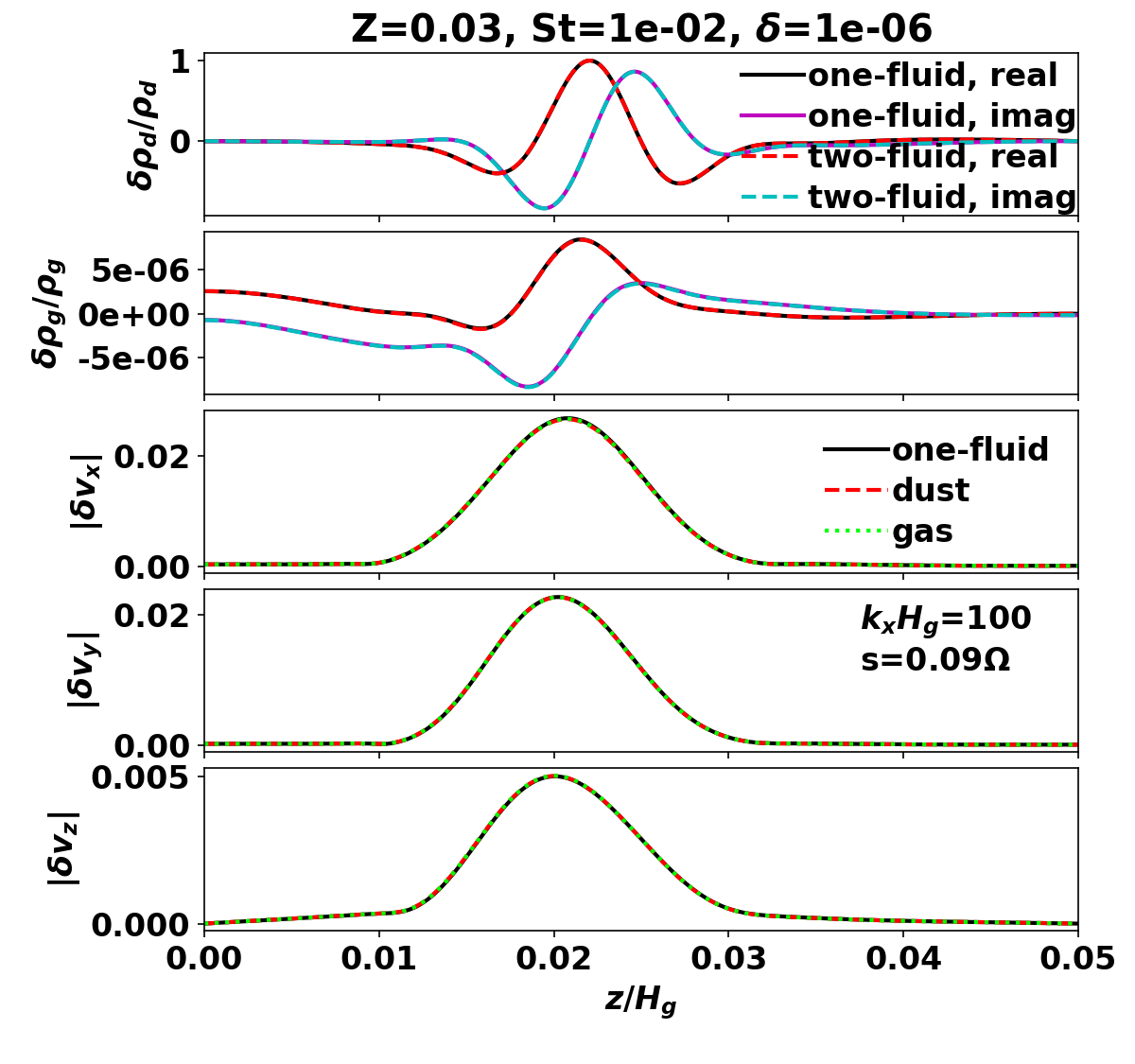}\includegraphics[width=.5\linewidth]{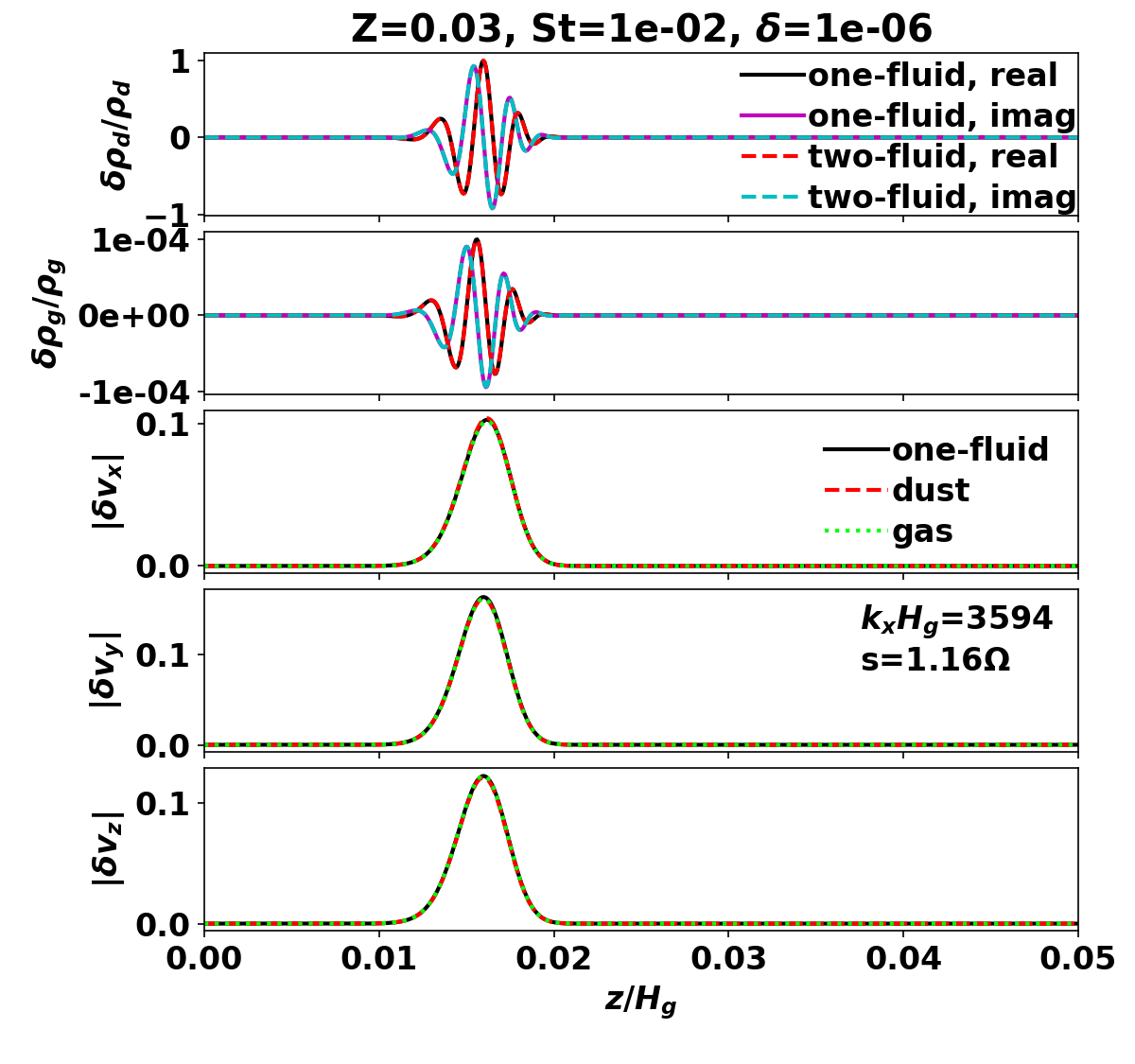}
    \caption{{Normalized eigenfunctions} of the most unstable modes found in case A {(high dust density layer)} with $K_x = 100$ (left) and {$K_x\simeq 3594$} (right, also the most unstable over $K_x$). Perturbations from top to bottom: relative dust density, relative gas density, radial velocity, azimuthal velocity, and vertical velocity. {For clarity we plot the amplitudes of the velocity eigenfunctions.}
    \label{caseA_eigenfunc}}
\end{figure*}

Example eigenfunctions from the above modes are shown in Fig. \ref{caseA_eigenfunc}. We find that with  increasing $K_x$, unstable modes become increasingly localized {about} $z\sim0.015\Hgas$. Notice there is little perturbation in the gas density for either mode, which indicate they are nearly incompressible.

\subsubsection{Pseudo-energy decomposition}
In order to identify physical origin of the above instabilities, we follow \cite{ishitsu09} and examine the energy-like quantity $U_\mathrm{tot} = \sum_{i=1}^{6}U_i$ associated with each mode, as described in Appendix \ref{2fluid_energy}. The contributions $U_i$
include: vertical shear in the equilibrium velocity field ($U_1$, which is dominated by the azimuthal component $U_{1y}$), vertical dust settling ($U_2$), pressure forces ($U_3$), dust-gas relative drift ($U_4$), {buoyancy ($U_5$), and viscosity ($U_6$). Note that $U_6\equiv0$ for inviscid disks, as considered here}.

\begin{figure*}
    \includegraphics[width=0.5\linewidth]{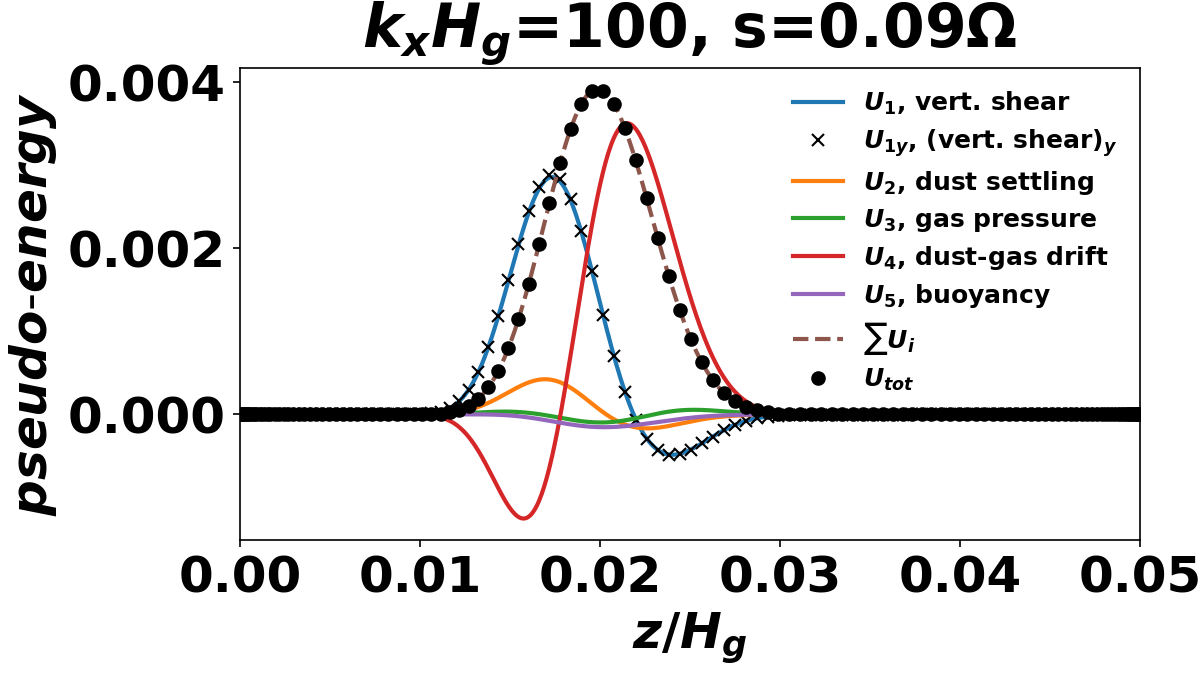}\includegraphics[width=0.5\linewidth]{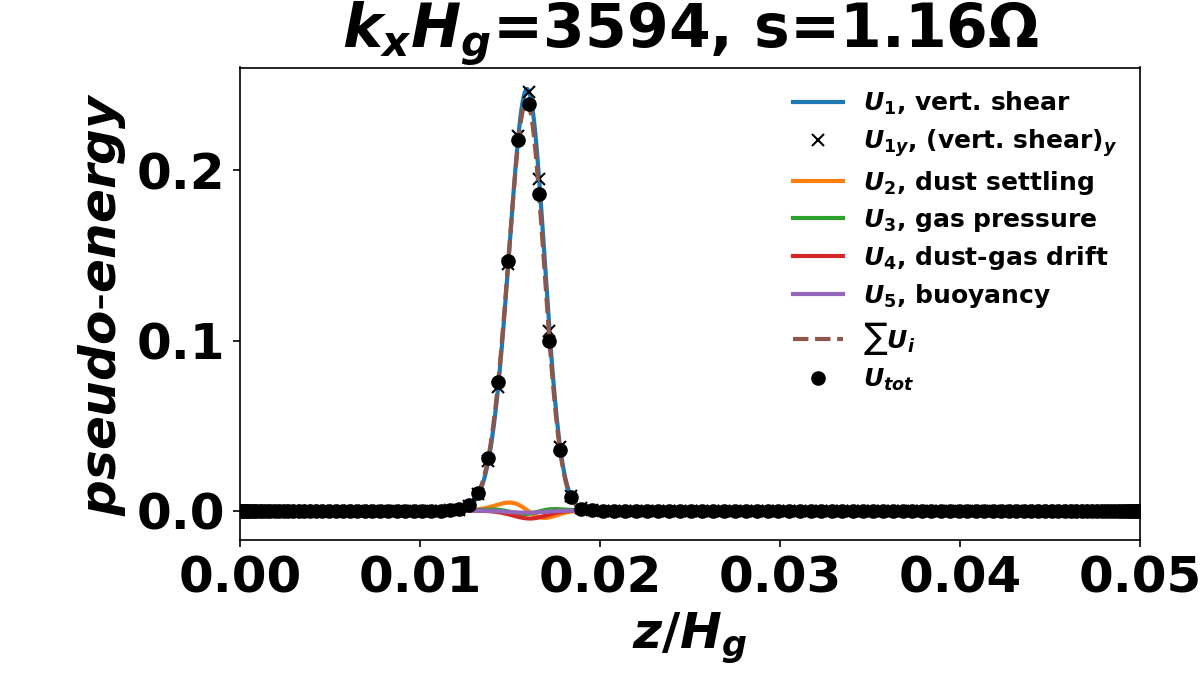}
    \caption{Pseudo-energy decomposition for the modes shown in Fig. \ref{caseA_eigenfunc}.
    \label{caseA_energy2f}}
\end{figure*}

Fig. \ref{caseA_energy2f} shows the pseudo-energy decomposition of the two main type of modes we find. For $K_x=100$ the mode is driven by a mixture of relative dust-gas drift ($U_4$, red), itself dominated by radial drift, and the vertical shear in the azimuthal velocity ($U_{1y}$, crosses). However, {the high $K_x=3594$} mode is entirely driven by vertical shear. 


\begin{figure*}
    \includegraphics[width=\linewidth]{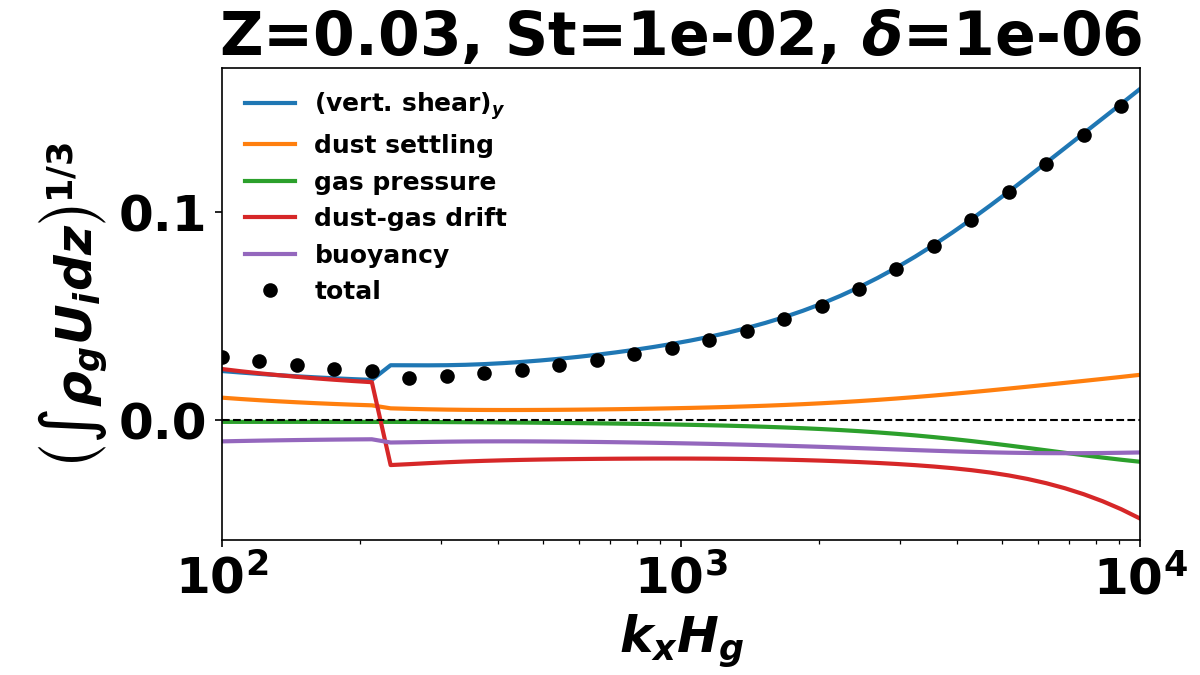}\caption{Vertically-integrated pseudo-energy contributions for the most unstable modes in case A, as a function of $K_x$. Note that we plot the cube root for improved visualization.
    \label{caseA_energy2f_int}}
\end{figure*}

In Fig. \ref{caseA_energy2f_int} we show the vertically-integrated pseudo-energy contributions as a function of $K_x$. We confirm that the abrupt change in oscillation frequencies around $K_x=200$ (see Fig. \ref{caseA_growth_max}) is due to a change in the character of the most unstable mode. For $K_x\lesssim 250$, modes are destabilized by a combination of dust-gas drift and vertical shear in the azimuthal velocity; while the latter dominates entirely for $K_x\gtrsim 250$. Interestingly, dust-gas drift becomes a stabilizing effect (its contribution becomes negative) for high-$K_x$ modes. On the other hand, vertical shear is always destabilizing.

Fig. \ref{caseA_energy2f_int} show that dust settling is always destabilizing, which is consistent with \cite{squire18b}, who find dust settling alone can lead to instability in vertically-local disk models. However, this effect is sub-dominant in our stratified models because vertical shear is much more significant. We also find that buoyancy forces are always stabilizing, as dust-gas coupling increases the mixture's inertia \citep{lin17}.

We find pressure forces provide increasing stabilization with increasing $K_x$, which is expected since pressure acts on small scales. {The increased restoring force from pressure may explain the increasing magnitude of oscillation frequencies  \citep{lubow93,balbus03}}. \cite{ishitsu09} showed that in the limit of {an} incompressible gas, pressure forces do not contribute to mode growth or decay. This indicates that gas compressibility becomes non-negligible for the high-$K_x$ modes in our case. These modes have short vertical wavelengths and are localized to regions of largest vertical shear (see Fig. \ref{caseA_eigenfunc}, right panel).{This is reminiscent of} `surface modes' of the gaseous VSI \citep{nelson13}, which are also stabilized by gas compressibility \citep{mcnally14}{. These similarities motivate us to interpret} the high-$K_x$, dust-driven, vertical-shear modes {as} dusty analogs of the gaseous VSI{, see \S\ref{dvsi}.}

\subsubsection{Dependence on the global pressure gradient}\label{pgrad_effect}

\begin{figure}
    \includegraphics[width=\linewidth]{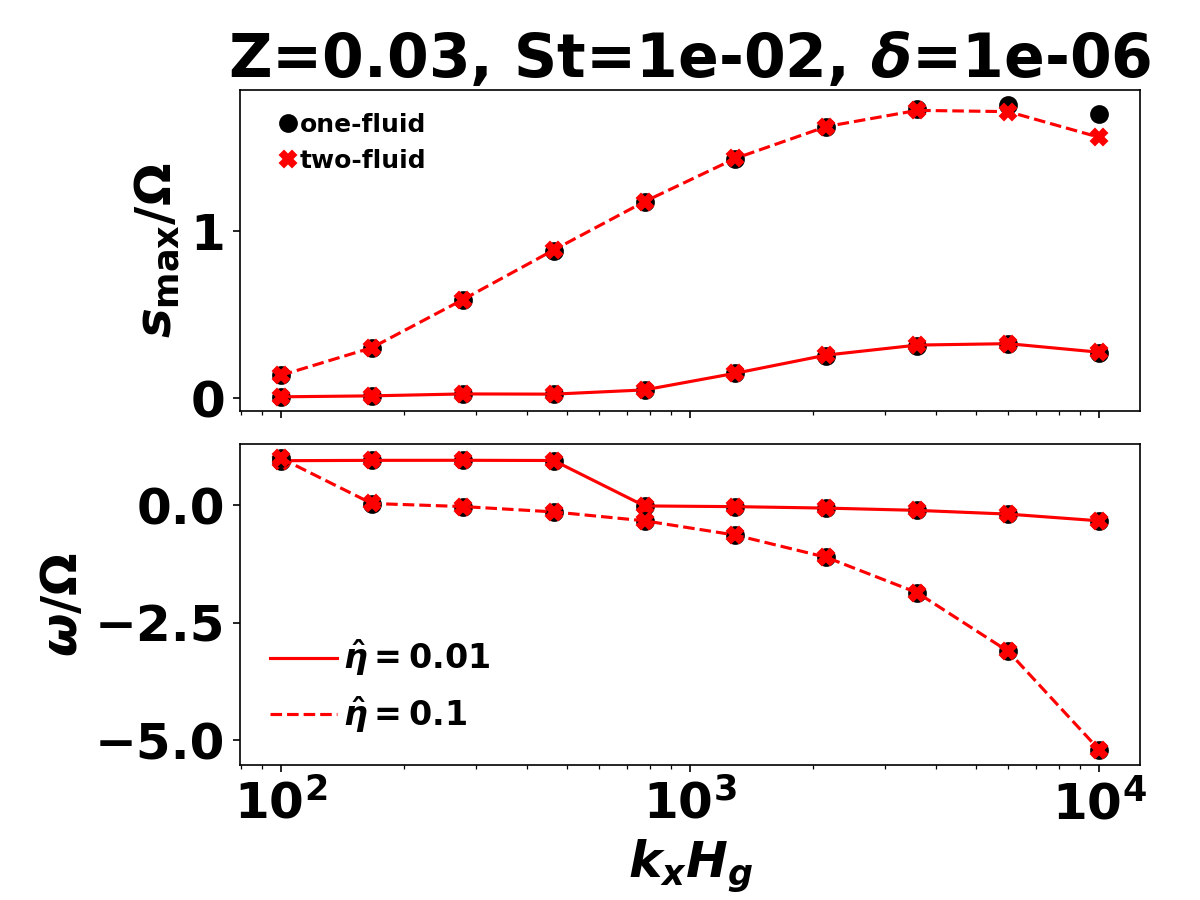}
    \caption{Maximum growth rate (top) and corresponding oscillation frequency (bottom) for unstable modes in case A {(high dust density layer)}, with different values of the radial pressure gradient $\hat{\eta}=0.01$ (solid) and $\hat{\eta}=0.1$ (dashed).
    \label{caseA_compare_eta}}
\end{figure}

Fig. \ref{caseA_compare_eta} shows the effect the global radial pressure gradient, as measured by $\hat{\eta} \equiv \eta r/\Hgas$. Note that dust-gas drift and vertical shear in the azimuthal velocity, which are the destabilizing effects for the modes we find, both scale with $\hat{\eta}$.
{This is consistent with the drift-driven classic SI, which also grows faster with increasing $\hat{\eta}$ at a fixed spatial scale \citep[][see their Eq. 29]{jacquet11}. 
For the vertical shear-driven unstable modes, the discussion in Appendix \ref{dvsi} also indicate that a minimum $\hat{\eta}$ is needed for instability (see Eq. \ref{dVSI_criterion}).
Hence, we find growth rates increase with $\hat{\eta}$.
}

We also find that the transition to modes purely driven by vertical shear occurs at smaller $K_x$ for larger $\hat{\eta}$: $K_x >100$ for $\hat{\eta}=0.1$ and $K_x > 500$ for $\hat{\eta}=0.01$. This is again similar to the gaseous VSI as a weaker vertical shear requires larger radial wavenumbers to destabilize \citep{latter18}.


\subsubsection{Dependence on particle size}\label{caseA_compare_st}
\begin{figure}
    \includegraphics[width=\linewidth]{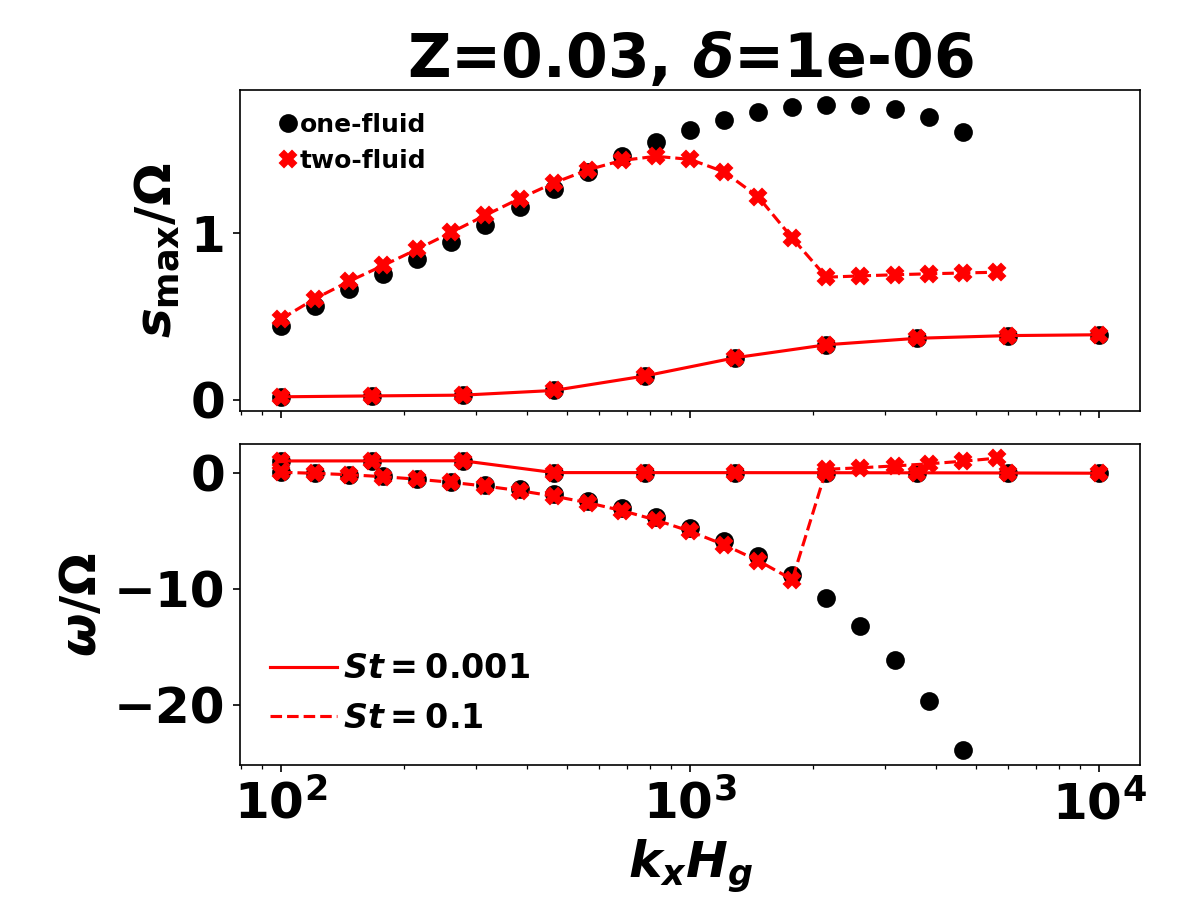}
    \caption{Maximum growth rate (top) and corresponding oscillation frequency (bottom) for unstable modes in case A {(high dust density layer)}, with different Stokes numbers: $\st=10^{-3}$ (solid) and $\st = 0.1$ (dashed).
    \label{caseA_compare_stokes}}
\end{figure}

Here we consider Stokes numbers $\st=10^{-3}$ and $\st=0.1$\footnote{In the two-fluid calculation with $\st = 0.1$, numerical artifacts developed near the disk surface, which were remedied by tapering the equilibrium vertical dust velocity to zero near the upper $10\%$ of the domain. However, this had negligible effects on the growth rates, oscillation frequencies, or the eigenfunctions in the disk bulk.}. The midplane dust-to-gas ratios are then $\sim 1$ and $\sim 9$, respectively {(see \S\ref{vert_eqm})}.
Growth rates and frequencies are shown in Fig. \ref{caseA_compare_stokes}. The trend for $\st=10^{-3}$ is {qualitatively} similar to our fiducial case with $\st=10^{-2}$, but with reduced growth rates. The modes again transition from drift-dominated to vertical-shear dominated as $K_x$ increases, here beyond $\sim 400$, somewhat higher than the fiducial case. {The oscillation frequencies for the $\st=10^{-3}$ vertical-shear modes are also much smaller in magnitude than that for $\st=10^{-2}$.}

{For $\st=0.1$ the curves in Fig. \ref{caseA_compare_stokes} are truncated at $K_x \gtrsim 5000$---$6000$ as we were unable to find converged two-fluid solutions; and one-fluid  eigenfunctions were found to have large, unphysical oscillations near the disk boundary. Notice also the one-fluid model over-predict growth rates for $K_x\gtrsim 10^{3}$. This is not surprising as the modes have $\st|\sigma|\gtrsim \Omega$, which can invalidate the one-fluid approximation \citep{lin17,paardekooper20}.

Nevertheless, we find $\st=0.1$ growth rates exceed that for $\st=10^{-3}$. Moreover, all the $\st=0.1$ modes are driven by vertical shear. However, for $K_x\gtrsim 2000$ the two-fluid modes have nearly constant growth rates and were found to be centered around $z=3.3\Hdust$, unlike the $K_x\lesssim 2000$ modes which are centered around $2\Hdust$, including the most unstable mode.
}

\begin{figure}
    \includegraphics[width=\linewidth]{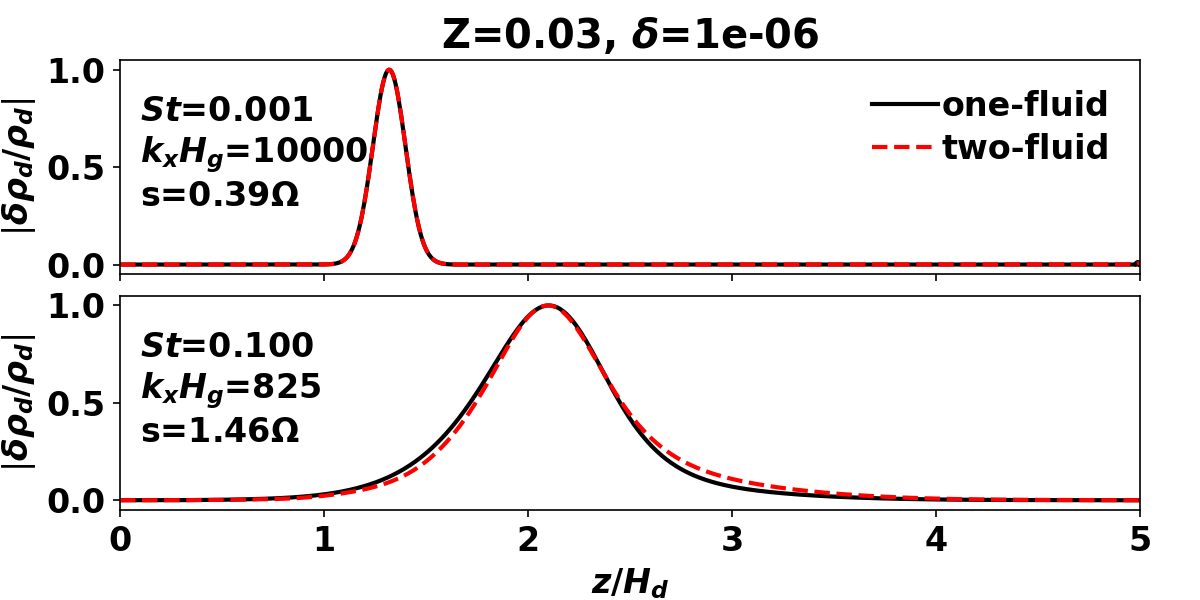}
    \caption{Magnitude of the relative dust density perturbation for the most unstable mode (over $K_x $, based on two-fluid growth rates) in case A with  $\st=10^{-3}$ (top) and $\st=0.1$ (bottom).
    \label{caseA_compare_stokes_eigenfunc}}
\end{figure}

The two-fluid results show that the most unstable modes in the $\st=10^{-3}$ and $\st=0.1$ disks occur at $K_x\gtrsim 10^{4}$ and $K_x\sim 825$, respectively. That is, instability with larger particles occur on larger radial scales. Fig.  \ref{caseA_compare_stokes_eigenfunc} compares the vertical profiles in the relative dust density perturbations for the most unstable modes. Here, we re-scale the vertical co-ordinate to account for different dust scale heights in these cases. For $\st=10^{-3}$ the mode is vertically-localized with a characteristic lengthscale  $l_z\simeq \Hdust/2$; whereas for $\st=0.1$ we find $l_z\simeq 2\Hdust$. Thus instability with larger particles are also more global in the vertical direction.



We interpret the above results as looser dust-gas coupling provides more rapid `cooling' to mitigate buoyancy forces, which then allows destabilization of disturbances on longer lengthscales,  similar to the gaseous VSI \citep[][see also \S\ref{dvsi}]{lin15}.


\subsubsection{Dependence on dust abundance}

\begin{figure}
    \includegraphics[width=\linewidth]{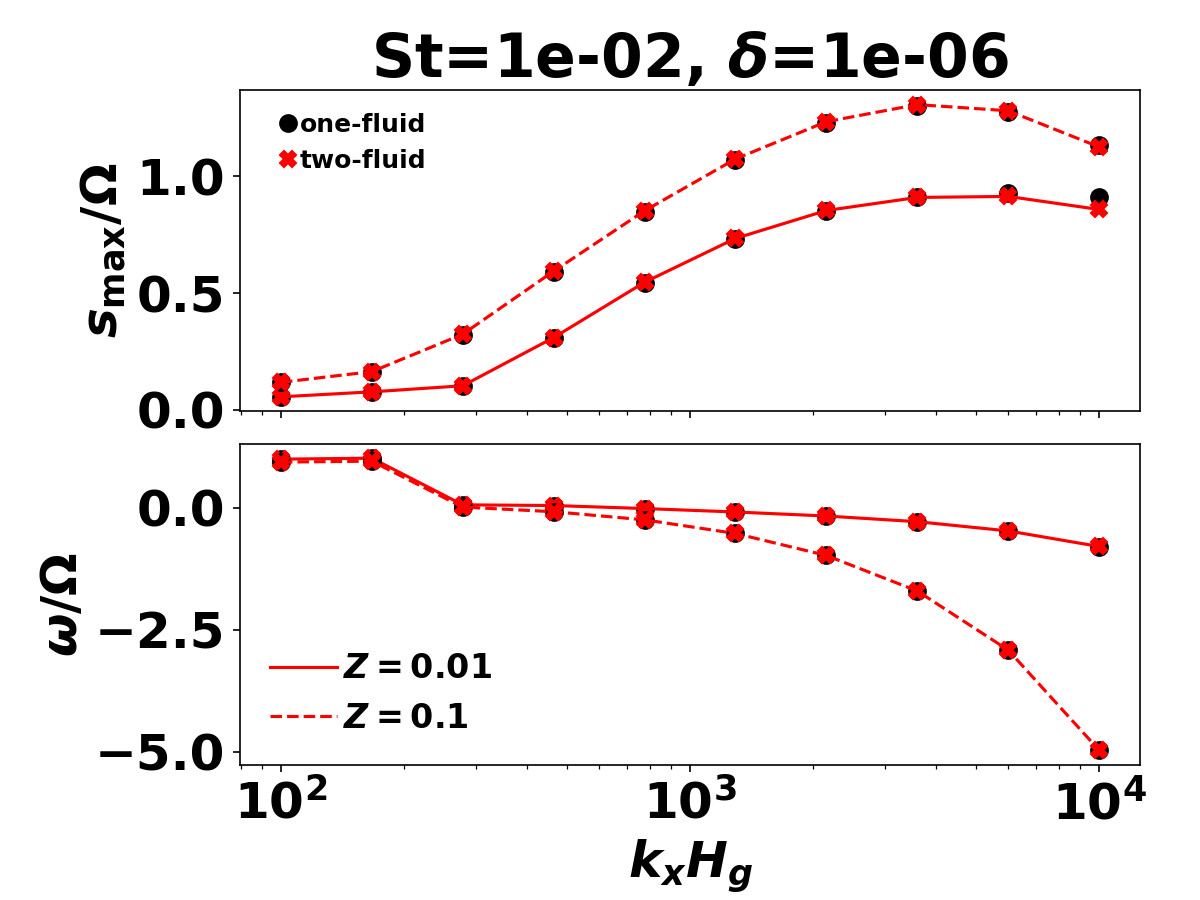}
    \caption{Maximum growth rate (top) and corresponding oscillation frequency (bottom) for unstable modes in case A {(high dust density layer)}, with different metallicities: $Z=0.01$ (solid) and $Z=0.1$ (dashed).
    \label{caseA_compare_Z}}
\end{figure}

In Fig. \ref{caseA_compare_Z} we plot the maximum growth rates and corresponding frequencies for modes in disks with different metallicities. We find growth rates are modestly increased with increasing solid abundance, but overall the results are insensitive to $Z$. In particular, the transition from mixed-modes to vertical shear-dominated modes does not depend on $Z$.

\subsection{Case B: {Low dust density layer}}\label{caseB}

\begin{figure}
    \includegraphics[width=\linewidth]{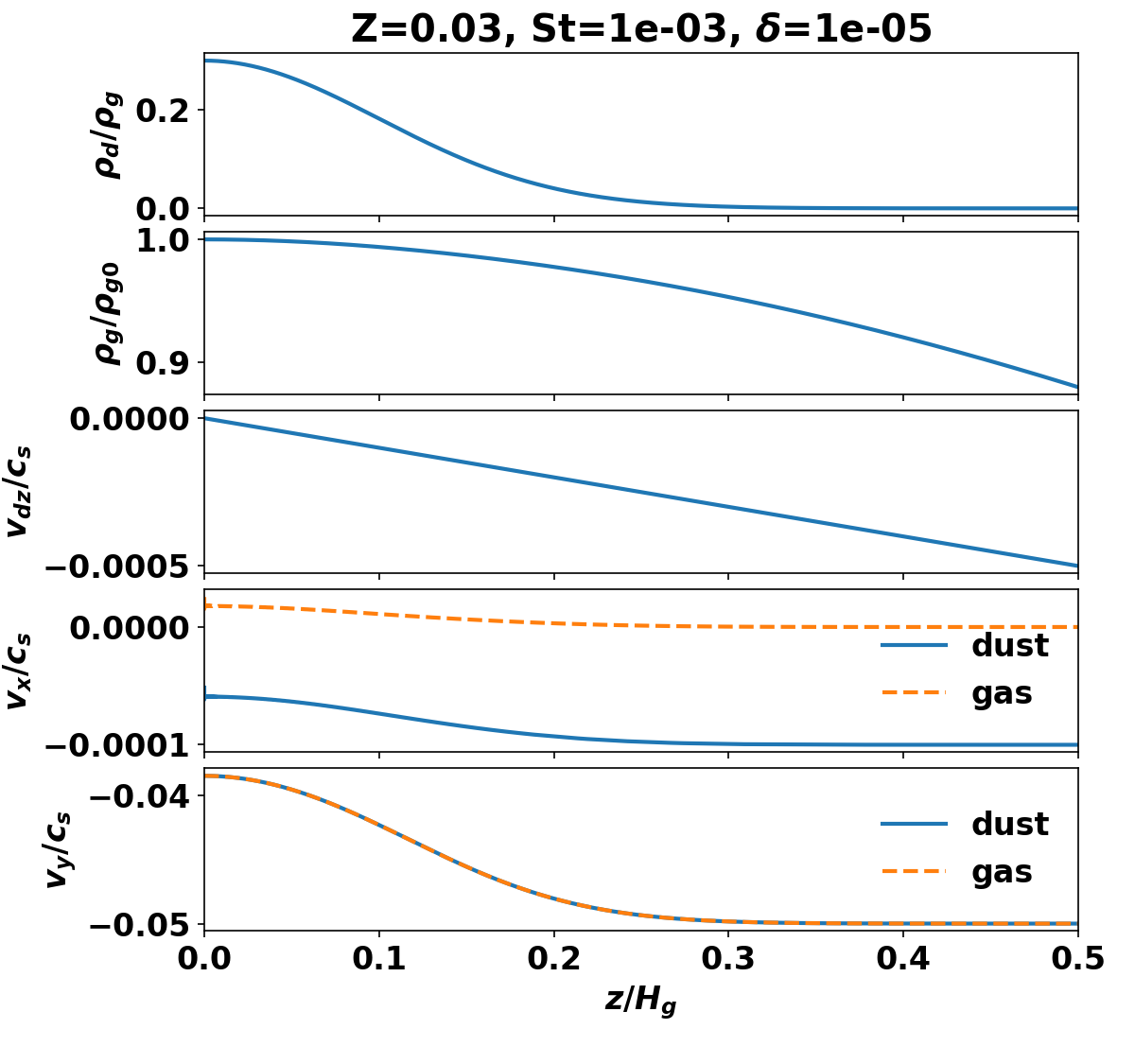}
    \caption{Two-fluid equilibrium for case B {with a low dust density layer}. From top to bottom: dust-to-gas ratio, gas density, vertical dust velocity, radial velocities, and azimuthal velocities.
    \label{caseB_eqm}}
\end{figure}

We now consider a {low dust density layer} with $\epsilon < 1$ throughout the disk column by choosing a stronger diffusion coefficient, $\delta \simeq 10^{-5}$, and smaller particles, $\st = 10^{-3}$. Other parameters are the same as the fiducial setup in case A. The equilibrium disk profile for case B is shown in Fig. \ref{caseB_eqm}.

\begin{figure}
    \includegraphics[width=\linewidth]{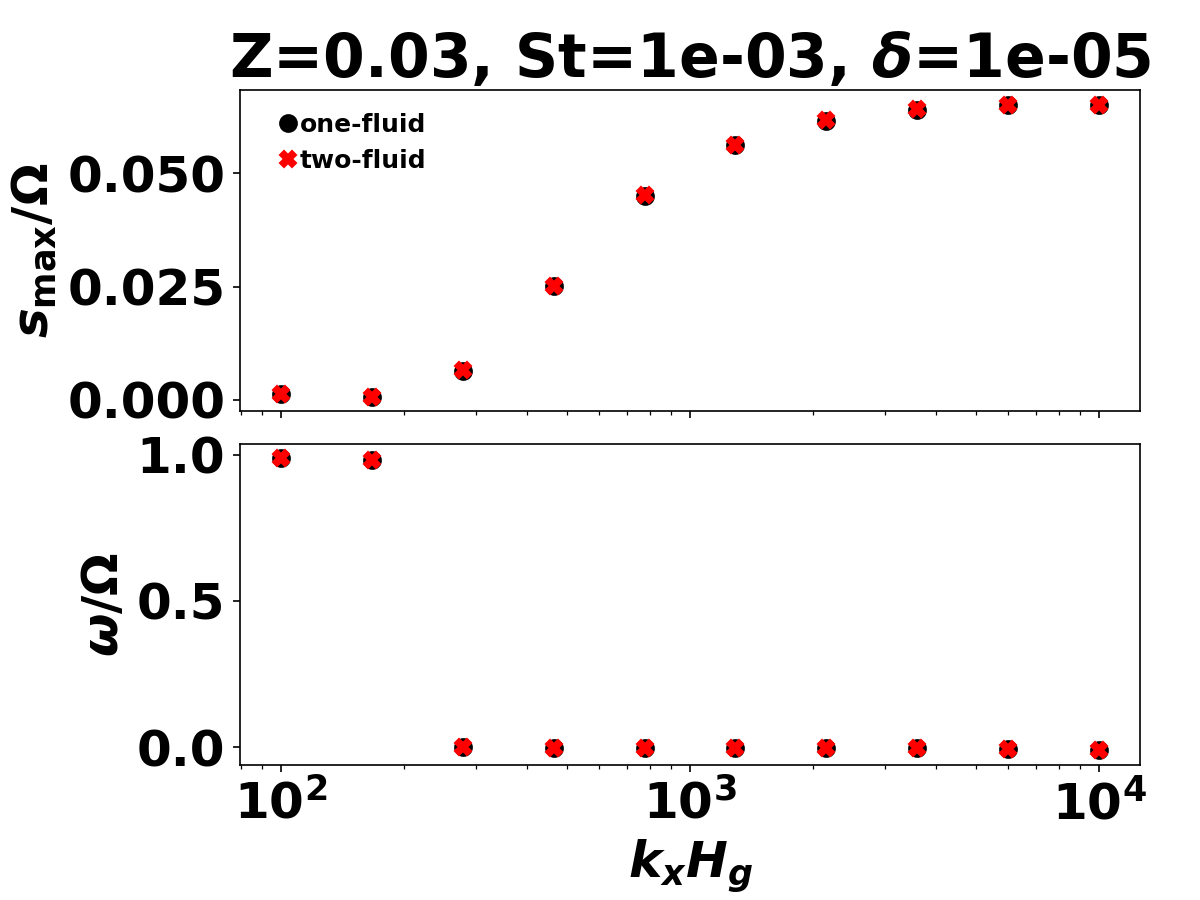}
    \caption{Maximum growth rate (top) and corresponding oscillation frequency (bottom) for unstable modes in case B {(low dust density layer)}, as a function of the dimensionless radial wavenumber $K_x$.
    \label{caseB_growth_max}}
\end{figure}

Growth rates and oscillation frequencies for case B are shown in Fig. \ref{caseB_growth_max}. We again find two distinct classes of unstable modes: for $K_x<200$ growth rates are small ($s\lesssim 10^{-2}\Omega$) with oscillation frequencies $O(\Omega)$; while for $K_x > 200$ modes are nearly purely growing with $s$ saturating around $0.06\Omega$. Case B is much more stable than case A owing to the smaller dust-to-gas ratio and particle size.

\begin{figure*}
    \includegraphics[width=\linewidth]{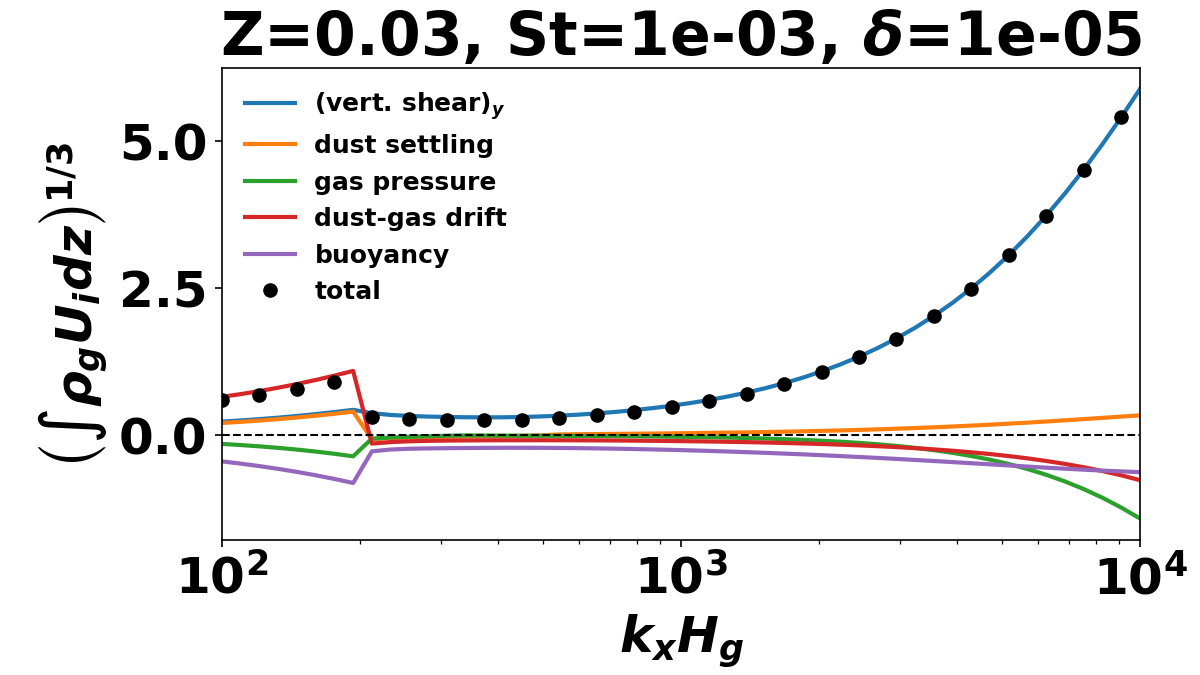}\caption{Vertically-integrated pseudo-energy contributions for most unstable modes in case B as a function of $K_x$. Note that we plot the cube root for improved visualization.
    \label{caseB_energy2f_int}}
\end{figure*}

Fig. \ref{caseB_energy2f_int} shows the vertically-integrated pseudo-energies. Modes with $K_x\lesssim 200$ are dominated by dust-gas drift with minor contributions from vertical shear and dust settling. This is unlike for case A where low-$K_x$ modes have equal contributions from dust-gas drift and vertical shear. However, for  $K_x\gtrsim 200$ modes are driven by vertical shear, as observed for case A.

\begin{figure}
    \includegraphics[width=\linewidth]{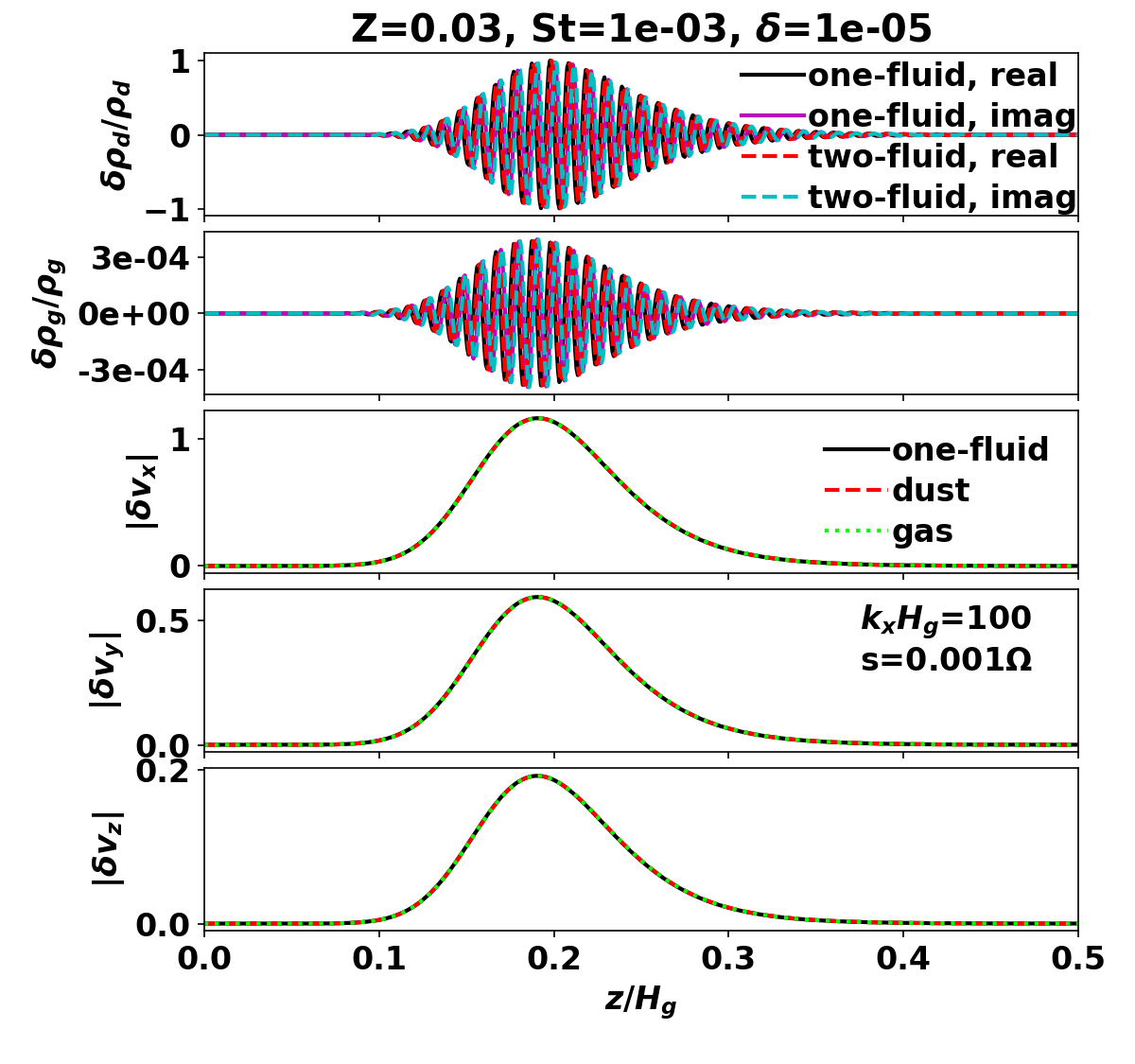}
    \caption{{Normalized eigenfunctions} of the most unstable mode found in case B {(low dust density layer)} with $K_x = 100$. Perturbations from top to bottom: relative dust density, relative gas density, radial velocity, azimuthal velocity, and vertical velocity. {For clarity, we plot the amplitudes of the velocity eigenfunctions.}
    \label{caseB_eigenfunc}}
\end{figure}

Fig. \ref{caseB_eigenfunc} shows the mode with $K_x=100$, primarily driven by dust-gas drift, involves ultra-short vertical oscillations {of characteristic lengthscale $10^{-2}\Hgas$, which is much smaller than $\Hdust\simeq 0.1\Hgas$}. This should be compared to the case A mode in the left panel of Fig. \ref{caseA_eigenfunc}, which is driven by a combination of relative dust-gas radial drift and vertical shear, and is more global{, i.e. it varies on a scale comparable to the dust layer thickness. This suggests that vertical-shear drives a more global disk response.} 

For {the high-$K_x$ modes driven mostly} by vertical shear we find similar behaviors in the eigenfunctions between case B and A: modes become increasingly localized with increasing $K_x$.

\subsection{Case C: Viscous disk}\label{caseC}

We briefly examine a viscous disk. To obtain appreciable growth rates in the presence of viscosity we set $\alpha = 10^{-7}$. {This value is much smaller than that expected in PPDs, but is sufficient to demonstrate the impact of viscosity. We discuss this issue further in \S\ref{gas_visc}. Here, we} use fiducial values of $\hat{\eta}=0.05$ and $\st=10^{-2}$, but set $Z=0.01$ so that the midplane dust-to-gas ratio $\epsilon_0\simeq 3$ is similar to case A. We also use a larger domain with $\zmax=7\Hdust$ as we find viscous modes at the smaller $K_x$ values tend to be vertically extended, a result already hinted by unstratified calculations \citep{chen20,umurhan20}.

\begin{figure}
    \includegraphics[width=\linewidth]{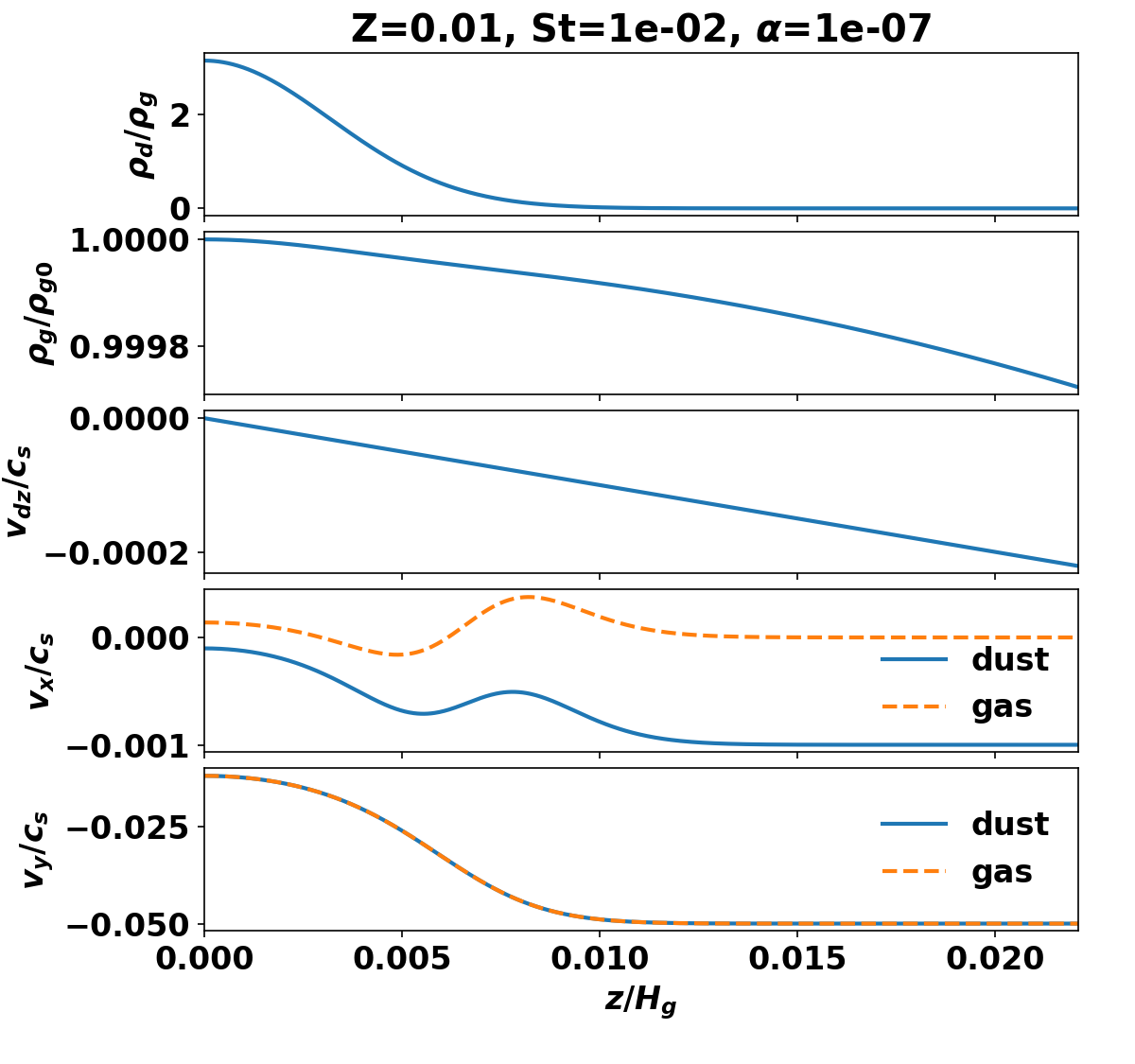}
    \caption{Two-fluid equilibrium for the viscous case C. From top to bottom: dust-to-gas ratio, gas density, vertical dust velocity, radial velocities, and azimuthal velocities.
    \label{caseC_eqm}}
\end{figure}

Fig. \ref{caseC_eqm} show that the equilibrium structure for this viscous disk is qualitatively similar to case A (Fig .\ref{caseA_eqm}), except in the radial velocities, which is noticeably non-monotonic away from the mid-plane.

\begin{figure}
    \centering
    \includegraphics[width=\linewidth]{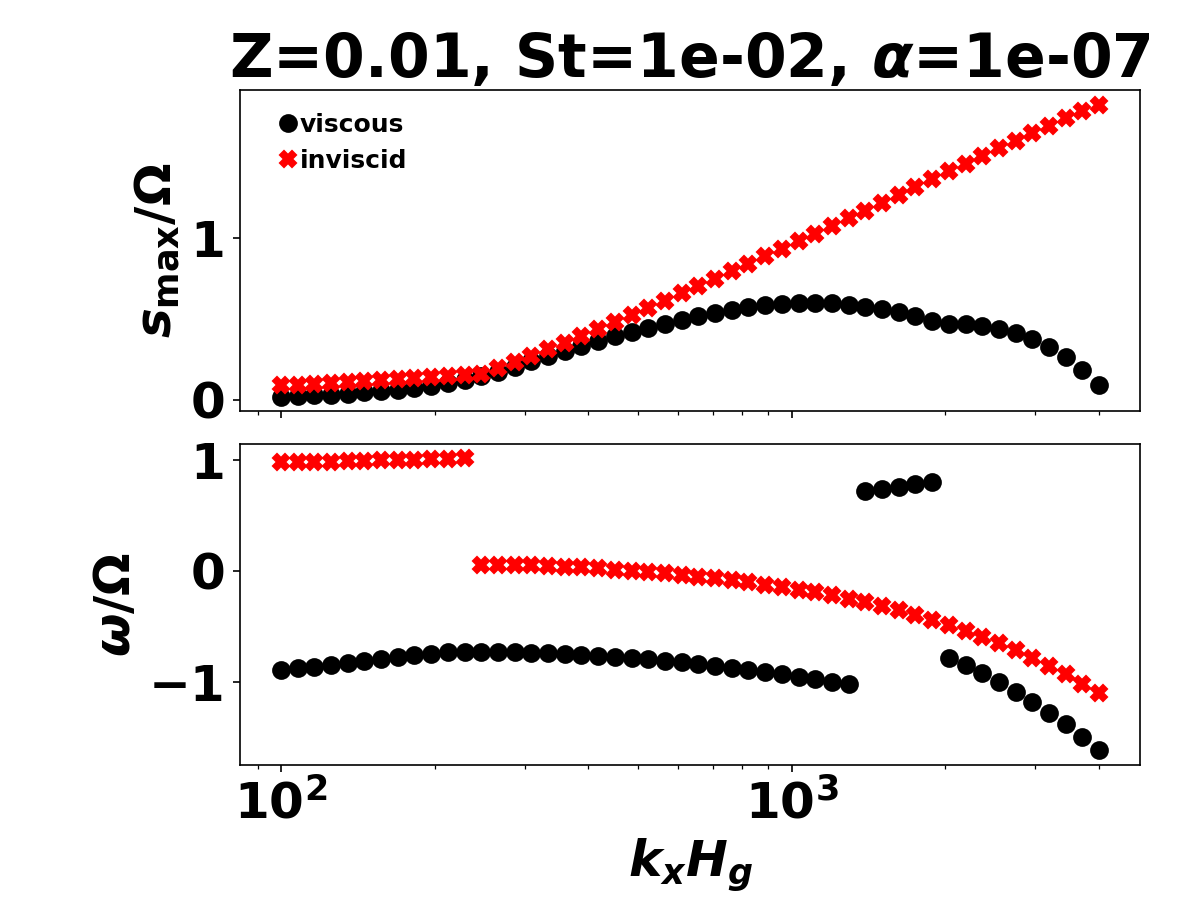}
    \caption{Growth rates (top) and oscillation frequencies (bottom) for modes found in the viscous case C (black circles) as a function of the dimensionless radial wavenumber $K_x$. Corresponding results for an inviscid disk are also shown (red crosses).
    \label{caseC_growth}}
\end{figure}

Fig. \ref{caseC_growth} shows the growth rates of the most unstable modes as a function of $K_x$ and corresponding oscillation frequencies. For comparison, we also plot results for an inviscid disk. We find viscosity strongly suppresses dust-gas instabilities. In the viscous disk, growth rates maximize at $K_x\simeq 1110$ with $s\sim 0.6\Omega$ and is essentially quenched for $K_x\gtrsim 4300$; while growth rates continue to increase with radial wavenumber in the inviscid disk.

\begin{figure*}
    \centering
    \includegraphics[width=\linewidth,clip=true,trim=0cm 2cm 0cm 2cm]{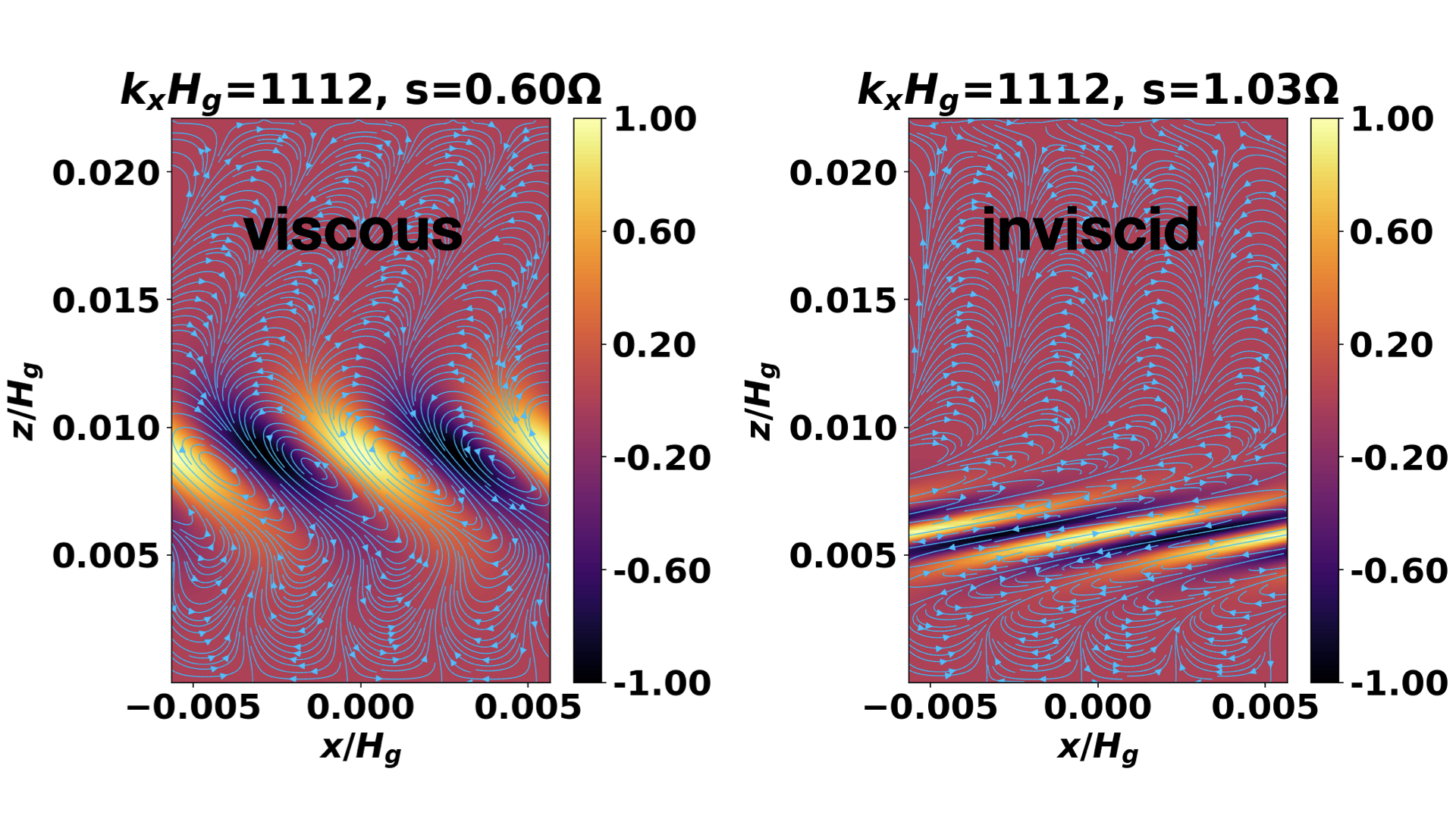}
    \caption{Structure of the most unstable mode found in the viscous case C (left, with $K_x\simeq 1110$) and the corresponding structure for an unstable {mode} in an inviscid disk with the same radial wavenumber. {Streamlines correspond to the perturbed dust velocity field and colors correspond to the relative dust density perturbation.}
    \label{caseC_eigenfunc_2D}}
\end{figure*}

\begin{figure}
    \centering
    \includegraphics[width=\linewidth]{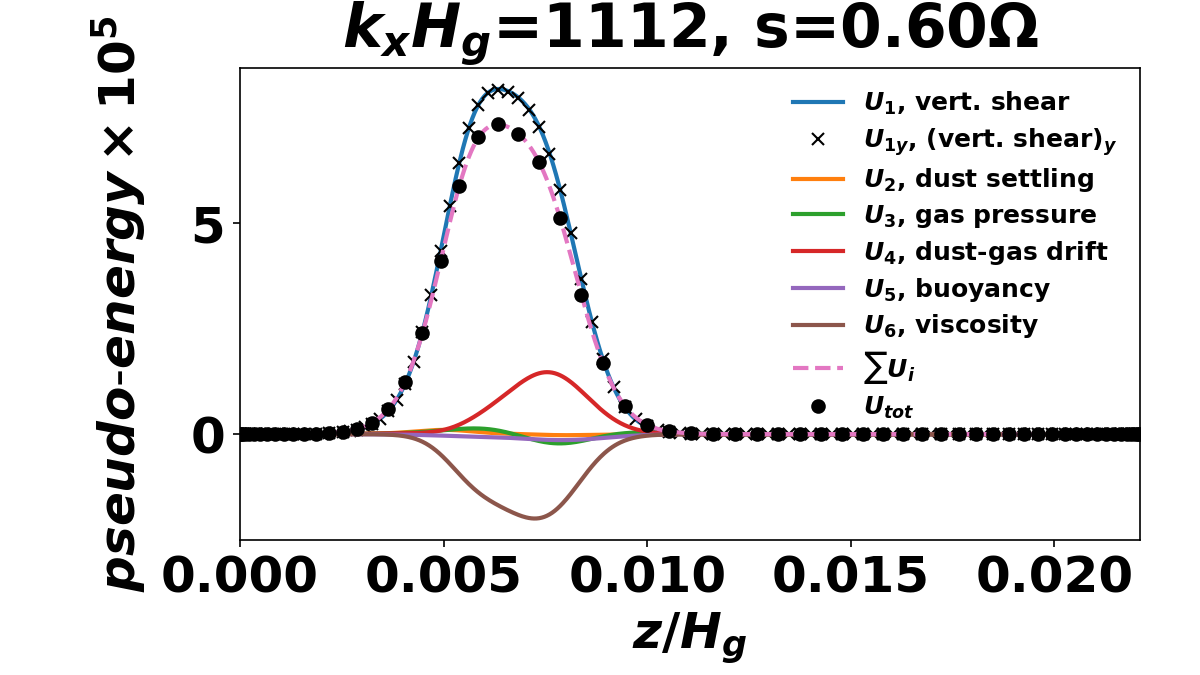}
    \caption{{Pseudo-energy decomposition of the most unstable mode found in the viscous disk (case C).}
    \label{caseC_energy2f}}
\end{figure}

In the left panel of Fig. \ref{caseC_eigenfunc_2D} we show a meridional
visualization of the most unstable mode found for case C. The corresponding flow in the inviscid disk is shown in the right panel. As expected, viscosity tends produce vertically-elongated disturbances{, here with length scales}  $\sim 1$-2$\Hdust$. {This mode is again predominantly driven by vertical shear, as demonstrated by its pseudo-energy decomposition shown in Fig. \ref{caseC_energy2f}.}

\begin{figure*}
    \centering
    \includegraphics[width=\linewidth]{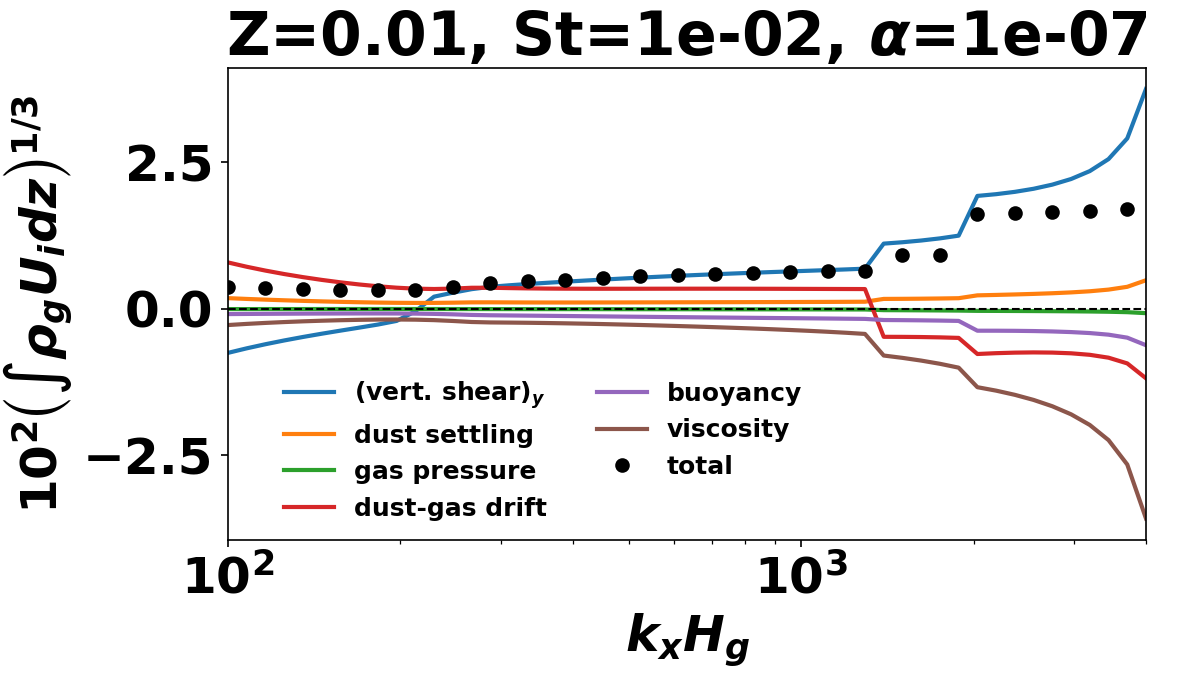}
    \caption{Vertically-integrated pseudo-energy contributions to the most unstable modes found in the viscous case C, as a function of $K_x$. Note that we take the cube root for improved visualization.
    \label{caseC_energy2f_int}}
\end{figure*}

Fig. \ref{caseC_energy2f_int} shows the vertically-integrated pseudo-energy contributions to the unstable modes. Note that there is now a viscous contribution (brown curve), see Appendix \ref{2fluid_energy}. For $K_x\lesssim 200$ unstable modes are driven by the relative dust-gas radial drift with minor contributions from dust settling. These modes have relatively small growth rates ($s\lesssim 0.1\Omega$). For $200\lesssim K_x\lesssim 1300$, modes are driven by vertical shear with  contributions from dust-gas drift. For $K_x\gtrsim 1300$, modes are {mostly} driven by vertical shear, but their growth rates decline rapidly due to viscosity, which is more effective at stabilizing smaller lengthscales.

As expected, buoyancy and viscous forces always act to stabilize the system. On the other hand, dust settling is always destabilizing \citep{squire18b}{, although here its effect is small.} Gas pressure has negligible effects, which reflects the incompressible nature of all the unstable modes presented. Like the inviscid cases A and B, we find dust-gas drift  becomes stabilizing at high radial wavenumbers (here $ \gtrsim 10^3$). However, unlike those inviscid cases where vertical shear is always destabilizing, in the viscous disk we find that for low $K_x$ modes ($\lesssim 200$) vertical shear becomes stabilizing.


\section{Discussion}{\label{discussion}}

\subsection{{Vertically-shearing streaming instabilities}}\label{dvsi}

Our numerical results show that the most unstable modes in stratified dust layers occur on radial lengthscales $\lesssim 10^{-3}\Hgas$ and are driven by the vertical gradient in the dusty disk's azimuthal velocity {combined with partial dust-gas coupling.} This is similar to the VSI in gaseous PPDs {\citep{nelson13}}. To interpret these {vertically-shearing streaming instabilities (VSSIs)}, we invoke the analogy between isothermal dusty gas and a pure gas subject to cooling as developed by \cite{lin17}.

For the gaseous VSI, the destabilizing vertical shear results from the global radial temperature gradient. However, in PPDs vertical {gas} buoyancy is strongly stabilizing. The gaseous VSI thus requires rapid cooling to remove the effect of buoyancy. In terms of these physical quantities, \cite{lin15} found the instability criterion

\begin{align}\label{vsi_condition}
    t_\mathrm{cool} < \frac{\left|\p_z v_y \right|}{N_z^2},
\end{align}
where $\p_z v_y$ is the vertical shear rate, $N_z$ is the vertical buoyancy frequency, and $t_\mathrm{cool}$ is the thermal cooling timescale such that the linearized cooling rate is
\begin{align}\label{tcool_gas_def}
    \delta \Lambda = - \frac{1}{\tcool}\left(\delta P - \frac{P}{\rhog}\delta \rhog \right).
\end{align}


We can obtain a criterion analogous to Eq. \ref{vsi_condition} for dusty disks as follows. We treat the isothermal dusty gas as a single fluid subject to a special cooling function, as described in Appendix \ref{one_fluid_model}. The dusty disk's azimuthal velocity profile is given by Eq. \ref{1fluid_eqm_vy_sol}. The vertical shear rate is thus
\begin{align}\label{vshear_dust}
\frac{\p v_y}{\p z} = \frac{\eta r \Omega \epsilon^\prime}{(1 + \epsilon)^2}.
\end{align}

Next, \cite{lin17} showed that the square of the vertical buoyancy frequency in a dusty disk is
\begin{align*}
    N_z^2 = c_s^2\frac{\p\ln{\rhog}}{\p z}\frac{\p \fdust}{\p z},
\end{align*}
where $\fdust = \epsilon/(1 + \epsilon)$ is the dust fraction. Using the equilibrium condition (Eq. \ref{1f_vz_eqm}) we find
\begin{align}\label{Nz2_dust}
    N_z^2 = -\frac{z \Omega^2\epsilon^\prime}{1+\epsilon}.
\end{align}

To estimate the appropriate `cooling time' in an isothermal dusty disk, we examine the one-fluid effective energy equation in Appendix \ref{one_fluid_model} (Eq. \ref{one_fluid_energy_linear}). We assume modes have small lengthscales and write $\p_z\to \ii k_z$, where $k_z$ is a real vertical wavenumber. Assuming both $k_x$ and $k_z$ are large in magnitude, the RHS of Eq. \ref{one_fluid_energy_linear}, which can be interpreted as a cooling rate after multiplying the equation by $P$, can be approximated by its leading term
\begin{align}\label{tcool_dust_motivation}
    \delta \Lambda_\mathrm{d} \equiv  -\frac{c_s^2\st\epsilon k^2}{\left(1 + \epsilon\right)^2\Omega}\delta P,
\end{align}
where $k^2 = k_x^2 + k_z^2$.
Comparing Eq. \ref{tcool_dust_motivation} with Eq. \ref{tcool_gas_def} motivates the identification
\begin{align}\label{tcool_dust_def}
    t_\mathrm{cool,d} \equiv \frac{(1+\epsilon)^2\Omega}{c_s^2\st \epsilon k^2} = \frac{(1+\epsilon)^2}{\epsilon \st K^2}\Omega^{-1}
\end{align}
as the cooling timescale of an isothermal dusty gas, where $K = k\Hgas$.

Inserting Eq. \ref{tcool_dust_def}, \ref{Nz2_dust}, and \ref{vshear_dust} into Eq. \ref{vsi_condition} gives the minimum wavenumber needed to trigger the dust-driven VSI,
\begin{align*}
    K^2 > \frac{(1+\epsilon)^3}{\epsilon \st \hat{\eta}}\left(\frac{\Hdust}{\Hgas}\right)\frac{z}{\Hdust}.
\end{align*}
Note that the RHS is a function of height. To obtain a more practical criterion, we evaluate it at $z = \Hdust$ and approximate 
$\epsilon \sim Z\Hgas/\Hdust$
(see Eq. \ref{metal_def}). For settled dust layers with $\st\gg \delta$, as considered throughout this work, we have $\Hdust/\Hgas \simeq \sqrt{\delta/\st}$. These approximations then give
\begin{align}\label{dVSI_criterion}
K^2 \gtrsim \frac{\left( 1 + Z\sqrt{\st/\delta}\right)^3\delta}{Z\hat{\eta}\st^2}.
\end{align}

Evaluating Eq. \ref{dVSI_criterion} for the fiducial case A ($Z=0.03$, $\st=10^{-2}$, $\delta = 10^{-6}$, $\hat{\eta}=0.05$) {suggest} $K\gtrsim 20$ is needed for finite dust-gas decoupling to destabilize the disk through vertical shear. Indeed, for case A we find vertical shear contributes to the instability for all $K_x$ considered ($\geq 100$), see Fig. \ref{caseA_energy2f_int}.


On the other hand, for case B we have $\st = 10^{-3}$ and $\delta\simeq 10^{-5}$, giving $K \gtrsim 120$. We thus expect a much larger $K_x$ is needed to tap into the free energy associated with vertical shear. Indeed, Fig. \ref{caseB_energy2f_int} show that vertical shear is sub-dominant for modes below the transition at $K_x\sim 200$. This is similar to the gaseous VSI \citep{lin15}: a slower `cooling' rate, here associated with stronger dust-gas coupling, means it is can only effectively destabilize smaller length scales.

{We remark that Eq. \ref{dVSI_criterion} can also applied to estimate the minimum radial pressure gradient needed to trigger VSSIs at a given spatial scale, which may explain the increasing growth rates with $\hat{\eta}$ seen in \S\ref{pgrad_effect}}.



\subsection{Comparison to \cite{ishitsu09}}
\cite{ishitsu09} carried out direct simulations to investigate the effect of a vertical density gradient on the stability {of} dust layers. Their disk models were initialized with a prescribed, non-uniform vertical dust density distribution and corresponding horizontal velocity profiles given by \cite{nakagawa86}. This is equivalent to stacking layers of unstratified disk models. They neglected vertical gravity, physical diffusion, viscosity, and assumed incompressible gas. We instead considered compressible gas (although this has negligible effects) and solve for the steady vertical disk structure self-consistently, which requires one to at least include dust diffusion.

Our results are broadly consistent with \citeauthor{ishitsu09}. Their simulation with $\st = 10^{-3}$ yield disturbances with characteristic wavenumber $k_x\eta r\sim 50$ (or $K_x \sim 10^3$ assuming $\hat{\eta}=0.05$) and growth rate $\sim \Omega$. This is comparable to our case A with $\st=10^{-3}$ in \S\ref{caseA_compare_st} (see Fig. \ref{caseA_compare_stokes}), for which we find growth rates of $0.2$--$0.4\Omega$ for $K_x\gtrsim 10^{3}$.
Importantly, \citeauthor{ishitsu09} also find that the vertical shear in the azimuthal velocity of the dust-gas mixture is the main driver of instability and that disturbances are centered off the disk midplane, similar to that observed in our {VSSI} eigenfunctions.


{
\subsection{Gas viscosity}\label{gas_visc}
The example in \S\ref{caseC} (case C) showed that even a small amount of gas viscosity of $\alpha = 10^{-7}$ can reduce growth rates significantly. However, in PPDs one may expect up to $\alpha\sim 10^{-4}$, for example due to turbulence driven by the gaseous VSI \citep{manger20}. Here, we discuss two supplementary calculations to further explore the role of gas viscosity.

We first repeat the fiducial case A, but enable viscosity ($\alpha=10^{-6}$) and use $\zmax=7\Hdust$. We find the most unstable mode, shown in Fig. \ref{gas_visc_2D_eigenfunc}, has $K_x\simeq 30$, $s \simeq 7\times10^{-4}\Omega$,  $\omega \simeq -3\times10^{-4}\Omega$, and is driven by vertical shear.  Unsurprisingly, the mode has a much larger spatial scale than that for case C (see Fig. \ref{caseC_eigenfunc_2D}), with a vertical lengthscale several times larger than $\Hdust$ ($=0.01\Hgas$).  However, the non-negligible mode amplitudes near $z=\zmax$ indicates that boundary conditions may be important.

\begin{figure}
    \includegraphics[width=\linewidth]{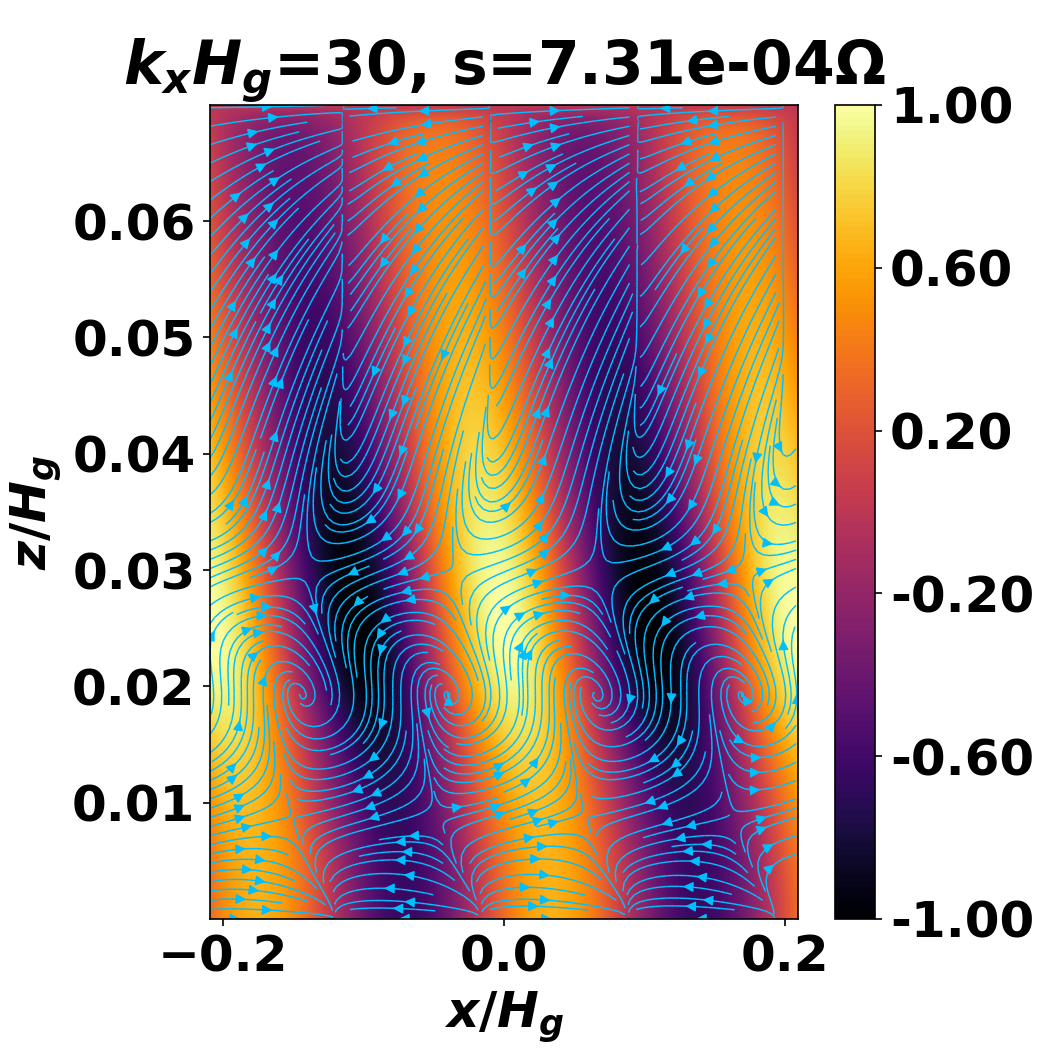}
    \caption{Structure of the most unstable mode in a viscous disk with $\alpha=10^{-6}$ and other disk parameters taken from case A (see \S\ref{caseA}).
    Streamlines correspond to the perturbed dust velocity field and colors correspond to the relative dust density perturbation.
    \label{gas_visc_2D_eigenfunc}}
\end{figure}

Next, we increase viscosity to $\alpha=10^{-4}$,  which alone would `puff up' the dust layer such that  $\epsilon<0.3$. We thus compensate by setting $Z=0.3$ to obtain the same mid-plane dust-to-gas ratio as the above case ($\epsilon_0\simeq 3$). We find the most unstable mode has $K_x\simeq 0.2$, $s\simeq 4\times10^{-4}\Omega$, and $s\simeq -\Omega$.
Interestingly, the mode is found to have comparable contributions from vertical shear and dust settling. However, its radial lengthscale of order $r$ (for $\hgas = 0.05$) calls the shearing box framework itself into question.


Nevertheless, these calculations are instructive to show that viscosity is strongly stabilizing and  produces disturbances that exceed the dust layer thickness, similar to the classic SI in unstratified, viscous disk models \citep{chen20,umurhan20}.
}

\subsection{Classic streaming instability}\label{classic_SI}
{
In unstratified disks the classic SI of \cite{youdin05a}, driven by the dust-gas relative radial drift, is usually the only linearly unstable mode \citep[but see][]{jaupart20},  which can lead to dust-clumping in the non-linear regime \citep{johansen07}.

In stratified disks, we find that the relative dust-gas drift can provide a significant (although not total) contribution to the most unstable modes on wavelengths of $O(10^{-2}\Hgas)$ or larger. Thus we still refer to them as classic SI modes. If their nonlinear  evolution is similar to their unstratified counterparts, then we can expect dust-clumping on such scales.

However, classic SI modes are not the most unstable across all scales, which are the VSSI modes that occur on radial scales of $O(10^{-3}\Hgas)$ or smaller, as discussed above. The nonlinear evolution of classic SI modes will thus be affected by small-scale VSSIs that develop first. If VSSIs saturate in turbulence, as the early work of  \citeauthor{ishitsu09} appear to suggest (albeit based on simulations with a limited parameter range and integration times -- see \S\ref{planetsimal_formation}) then we expect a reduced efficiency of dust-clumping via classic SIs.

In this case, we suggest that low-resolution (e.g. global)  simulations that are biased towards classic SI modes should include a physical dust diffusion to mimic the effect of unresolved VSSI turbulence.
}

\subsection{Dust settling instability}
The dust settling instability \citep[DSI,][]{squire18b,zhuravlev19,zhuravlev20} is an analog of the classic SI, except the DSI is driven by the vertical drift of dust relative to the gas (i.e. dust settling), rather than their relative radial drift. The DSI has been proposed to seed planetesimal formation by acting as a dust-clumping mechanism, although recent simulations show this effect may be weak in practice \citep{krapp20}. Studies of the DSI have so far adopted vertically-local disk models, which neglect  vertical shear, but permit equilibria to be defined without dust diffusion. This is not possible in our vertically-global disk models.

Our vertically-global models confirm that dust settling acts to destabilize stratified dusty disks, which suggest that the DSI is present. However, it is generally sub-dominant to vertical shear,  relative radial drift, or both. We can crudely understand this by comparing the dust settling velocity $\vdz$ to the azimuthal velocity difference across the dust layer, $\Hdust v_y^\prime$. Using Eqs. \ref{2f_eps}, \ref{2f_vdz}, \ref{vshear_dust}, and considering $\st\ll 1$, we find
\begin{align}\label{vshear_vs_settle}
\left|\frac{\Hdust v_y^\prime}{\vdz}\right|
\simeq \frac{\epsilon}{(1+\epsilon)^2} \frac{\hat{\eta}}{\delta}\frac{\Hdust}{\Hgas}
 \sim \frac{Z\hat{\eta}}{\left(1 + \epsilon\right)^2\delta},
\end{align}
where we used $Z\sim \epsilon \Hdust/\Hgas$ (see Eq. \ref{metal_def}).
{We can further use $\delta\simeq \st Z^2/\epsilon^2$ to rewrite Eq. \ref{vshear_vs_settle} as $\sim \hat{\eta}/(\st Z)$, assuming $\epsilon\gtrsim 1$.}


For the fiducial case A we find {Eq. \ref{vshear_vs_settle}  gives} $\sim 90$ i.e. 
dust settling is dwarfed by vertical shear. Hence for well-settled dust (small $\delta$), instability is primarily due to vertical shear, provided its free energy can be accessed (see \S\ref{dvsi}).

Dust settling may dominate over vertical shear if $\delta \gtrsim Z\hat{\eta}$. For PPDs with $Z$ and $\hat{\eta}$ both of $O(10^{-2})$ this requirement translates to $\delta \gtrsim 10^{-4}$. However, such a large diffusion parameter {is  likely to be strongly stabilizing} \citep{chen20,umurhan20,krapp20}.

Similarly, we can compare settling to the dust-gas relative radial drift (Eq. \ref{vdrift}), say at $z=\Hdust$, to find
\begin{align}
    \left|\frac{v_\text{drift}}{\vdz}\right|\sim \frac{2\epsilon}{1 + \epsilon}\frac{\hat{\eta}}{Z},
\end{align}
{again assuming $\st\ll 1$}.
In PPDs the last factor is $O(1)$. Then for settled dust layers with $\epsilon$ of $O(1)$ we expect settling to be at most comparable to radial drift.
For {well-mixed} dust layers {with $\epsilon \sim Z \sim O(10^{-2})$, the above ratio is $O(\hat{\eta})\ll 1$} so that dust settling can dominate.  However, having such an equilibrium requires a large diffusion coefficient ($\delta\gg \st$), which {may provide} complete stabilization.


{

\subsection{Applicability of RDI theory}
Both the classic SI (for $\epsilon \ll 1$) and the DSI are  `resonant drag instabilities'  \citep[RDI][]{squire18a,squire18b}. RDIs arise when the background relative dust-gas motion resonates with a wave in the gas. The local condition for an RDI is
\begin{align}\label{RDI_cond}
    \bm{k}\cdot\left(\vd - \vg\right) = \omega_\mathrm{gas}\left(\bm{k}\right)
\end{align}
\citep{squire18a}, where $\omega_\mathrm{gas}(\bm{k})$ is the frequency of a wave mode in the gas (when there is no dust) with local wavenumber $\bm{k}$. The small-$\epsilon$ classic SI and the DSI occur when  radial dust drift and vertical dust settling resonates  with inertial waves in the gas, respectively. It is then natural to ask whether or not vertically-global modes in our stratified disks can be also interpreted as RDIs.

The first step of the RDI recipe given by \cite{squire18b} is to choose a gas mode in the absence of dust. Fortunately, analytic dispersion relations can be obtained for stratified gas disks \citep{lubow93,lin15}. Specifically, inertial waves in an isothermal Keplerian disk satisfy
\begin{align}
    \omega_\mathrm{gas}^2 = \frac{L}{K_x^2 + L}\Omega^2,
\end{align}
where $L$ is an integer and it is assumed that $L\gg\omega_\mathrm{gas}^2$ \citep{barker15,lin15}.

Let us consider inertial waves with $L\gg K_x^2$. Then $\omega_\mathrm{gas}\simeq \Omega$, as observed in the oscillation frequency for modes with $K_x=100$ in cases A and B (see Figs. \ref{caseA_growth_max} and \ref{caseB_growth_max},  respectively). Using the radial drift and dust settling velocities given by Eq. \ref{vdrift} and \ref{2f_vdz}, respectively, Eq. \ref{RDI_cond} becomes
\begin{align*}
    -2\hat{\eta}K_x - \frac{z}{\Hgas}K_z = \frac{1}{\st},
\end{align*}
where we have assumed $\st, \epsilon\ll 1$ and $K_z = k_z\Hgas$. We can use this condition to estimate the vertical wavenumber $K_z$ required for resonance.

Consider the $K_x=100$ mode in case A with $\st=10^{-2}$ and $z=0.02\Hgas$ or case B with $\st=10^{-3}$ and $z=0.2\Hgas$, where radial drift and dust settling contributes to instability. The heights are chosen where mode amplitudes maximize, see Figs. \ref{caseA_eigenfunc} (left panel) and \ref{caseB_eigenfunc}. We then find $\left|K_z\right|\sim 5000$. However, the actual global   eigenfunctions are better characterized by $\left|K_z\right|\sim 10^{2}$, indicating such modes do not reflect a RDI, at least locally.

This discrepancy may be related to the fact that these modes are not purely associated with radial drift and dust settling: vertical shear also contributes (see Figs. \ref{caseA_energy2f_int} and   \ref{caseB_energy2f_int}), but this effect does not enter local RDI theory.

On the other hand, the dominant VSSI modes are unrelated to the relative dust-gas motion in the background disk. Instead, it is associated with the single azimuthal velocity of the dust-plus-gas disk. It is thus unclear if VSSI modes can be interpreted as RDIs.

It will be necessary to develop a global RDI theory to address the above issues.
}

\subsection{Implications for planetesimal formation}\label{planetsimal_formation}


In non-linear simulations, \cite{ishitsu09} showed that {VSSI modes first} lead to turbulence. They found large grains with unit $\st$\footnote{This regime cannot be {probed in our disk models} because no equilibrium can be defined, see \S\ref{vert_eqm}.} then underwent clumping, possibly due to the classic SI, which is most effective for marginally-coupled solids \citep{youdin05a}. {However}, small grains ($\st = 10^{-3}$) were dispersed by the turbulence and did not clump, but this may be due to insufficient metallicities and integration times.

Recent simulations carried out by \cite{yang17} show that the clumping of small grains ($\st=10^{-3}$--$10^{-2}$) require sufficient metallicities ($Z\gtrsim 0.02$--$0.04$) and integration times ($\gtrsim10^2$--$10^3$ orbits). They also found dust-clumping occurs after the disk saturates in a turbulent state. It is worth noting that the resolutions adopted in their simulations, of $O(10^{-4}\Hgas)$, should resolve {VSSI modes, which we find to dominate on radial scales of $O(10^{-3}\Hgas)$}.

Similarly, \cite{bai10} carried out three-dimensional simulations of stratified dusty disks, but found that the initial turbulence is largely axisymmetric,
which rules out non-axisymmetric KHIs as the cause  \citep{chiang08,lee10}. Given the large growth rates and axisymmetric nature of {VSSIs}, we suggest these were in fact responsible for the initial turbulence observed by {\citeauthor{bai10} and \citeauthor{yang17}}.


Similar to the gaseous VSI, {VSSI} turbulence is expected to erase the vertical shear responsible for it \citep{barker15}, i.e. dust stratification, by vertically mixing up solids \citep{stoll16,flock17,lin19}. Afterwards, we expect classic SI modes to become dominant. However, the ambient 
{VSSI} turbulence likely provides significant stabilization, especially for small grains \citep{chen20,umurhan20}. {This suggests that
planetesimal formation in PPDs may be less efficient than estimates based on unstratified, laminar disk {models}, because VSSI turbulence should always be present in realistically stratified disks.}

In light of the above discussion, we {hypothesize}  the following interpretation of planetesimal formation as observed in previous {simulations}, e.g. \cite{johansen09}. A thin, stratified dust layer first undergoes {VSSIs}. This leads to turbulence that renders the dust layer with {almost uniform density  and marginally stable against VSSIs, but still unstable to classic SI modes.}
The {nonlinear evolution of the classic SI then produce} dust clumps that, under appropriate conditions, leads to  gravitational collapse into planetesimals.

\subsection{Caveats and outlooks}\label{caveats}

\subsubsection{{Analytical models}}

We have relied on full numerical solutions to the linearized equations. Although our subsequent analyses hint at the physical origin of the various instabilities uncovered, a true understanding of the instability mechanisms require more rigorous mathematical modeling  \citep{jacquet11,squire18b,jaupart20,pan20a,pan20b}.

To this end, it is desirable to derive {an algebraic dispersion relation} for {modes in a} stratified {dusty} disk. {This will allow us to classify modes and explain their growth as well as oscillation frequencies.} This might be possible for the {VSSI} by {exploiting} the analogy between dust-laden flows and pure gas subject to cooling \citep{lin17}, as in the latter case analytic solutions for the gaseous VSI can be obtained \citep{lin15}.


{
\subsubsection{Multiple dust species}

We have only considered a single dust species. However, a distribution of particle sizes is expected in reality \citep{mathis77,birnstiel12}. Recent generalizations of the classic SI in unstratified disks show that having multiple dust species can significantly reduce growth rates when $\epsilon \lesssim 1$   \citep{krapp19,zhu20,paardekooper20}. For the DSI, though, a particle size distribution has a limited effect \citep{krapp20}.

The VSSI is associated with the vertical shear of the  dust-plus-gas system. Considering small, tightly coupled grains, all dust species and the gas share the same azimuthal velocity to $O(\st)$. We may thus naively expect the VSSI to be qualitatively similar for single and multiple dust species, if the two systems have the same dust-to-gas ratio profile and average Stokes number.
This should be checked with explicit calculations.


One approach is to add vertical gravity and dust diffusion to the one-fluid model of a dusty gas with a continuous particle size distribution recently developed by \cite{paardekooper20}. This is equivalent to adding one extra equation for the particle size-density to those presented in Appendix \ref{one_fluid_model}, which can then be implemented in the codes developed for this study.
}

\subsubsection{{Dust diffusion model}}\label{diff_model_problems}

We adopted a simple dust diffusion model so that stratified equilibria can be defined and standard linear stability analyses can be carried out. Physically, this model assumes there exists an underlying, external mechanism that stirs up dust grains, such as turbulence. In our implementation this is characterized by a single, constant diffusion coefficient. However, realistic turbulence may depend on the disk structure. For example, turbulence driven by the gaseous VSI can be reduced by dust-loading \citep{lin19,schafer20}. In this case, particle diffusion within the dusty midplane should be weaker than the dust-free gas above and below it.

Particle stirring may also result from dust-gas instabilities itself.  Consider, for example, an initially laminar disk. As grains settle, it may (instantaneously) meet the conditions for the classic SI, dust-driven VSI, KHIs, or others. However, to properly describe how a settling dust layer becomes unstable, one needs to perform stability analyses with respect to non-steady backgrounds \citep[e.g.][]{garaud04}, which is beyond the scope of this work. Nevertheless, such instabilities are thought to drive turbulence that prevents further settling \citep[e.g.][]{johansen09} and maintain a quasi-steady state.

In the above contexts, our study should be interpreted as the stability of dust layers whose equilibrium state is maintained by turbulence driven by pre-existing dust-gas instabilities. To better reflect realistic PPDs, our models should thus be generalized to diffusion (and possibly gas viscosity) coefficients with strength and spatial dependencies based on explicit simulations of quasi-steady, turbulent dust layers \citep[e.g.][]{bai10,yang17}.



{A more fundamental issue with the adopted diffusion model, though common, is that it can lead to non-conservation of total angular momentum \citep{tominaga19}. However, we suspect this will not to qualitatively affect the VSSI as it is unrelated to dust diffusion: \citeauthor{ishitsu09} observe the same instabilities, but only included a small diffusion term for numerical stability.  Nevertheless, it would be useful to examine the VSSI in the angular momentum-conserving formalism introduced by \citeauthor{tominaga19}.
}

\subsubsection{{Non-axisymmetry}}

Finally, we considered axisymmetric perturbations exclusively, which preclude non-axisymmetric KHIs \citep{chiang08,lee10}.
The growth of KHIs and the axisymmetric instabilities presented in this work should be compared to assess which is more relevant in PPDs. However, in the shearing box framework, linear, non-axisymmetric disturbances may only undergo transient or algebraic growth  \citep[e.g.][]{balbus92,johnson05}. Describing them requires one to solve an initial value problem, rather than the eigenvalue problem herein. Alternatively, one can forgo the plane wave ansatz and compute the full radial structure of linear disturbances with non-periodic boundary conditions \citep[e.g.][]{adams89,savonije90,lin11a,lin11b}. However, this would result in a partial differential equation eigenvalue problem \citep[e.g.][]{lin13}, which is significantly more complex than that considered in this work .


\section{Summary}\label{summary}
In this paper, we study the axisymmetric linear stability of vertically stratified dust layers in protoplanetary disks {(PPDs)}.
{Our disk models extend those used to study the classic streaming instability \citep[SI,][]{youdin05a}, namely unstratified disks, by accounting for the vertical structure of dust and gas in PPDs, as solids are expected to settle near the disk midplane.}

We find the dominant instability in stratified disks is one driven by the vertical gradient in the dusty-gas' azimuthal velocity.
 The large vertical shear within a settled dust layer is a significant source of free energy, which can be accessed via {partial} dust-gas coupling. This allows unstable modes to grow on orbital timescales. Our findings are consistent with earlier non-linear simulations carried out by \cite{ishitsu09}.

In PPDs, these {vertically-shearing streaming instabilities (VSSIs)} occur on radial scales  $\lesssim 10^{-3}\Hgas$, where $\Hgas$ is the local gas scale height. On the other hand, {classic SI modes}, associated with the relative radial drift between dust and gas, occur on radial length scales $\gtrsim 10^{-2}\Hgas$, but have {much} smaller growth rates than {VSSIs}.

However, the non-linear evolution of {VSSIs may drive} turbulence that mixes up the dust layer \citep{ishitsu09}{, rather than dust clumping like the classic SI \citep{johansen07}}. Given their dynamical growth rates, {we suggest VSSI turbulence may have} already manifested in {some  simulations} \citep[e.g.][]{bai10,yang17}{, which show that stratified dust layers first settle into a quasi-steady, turbulent state before clumping.

If VSSIs are inherent to PPDs and its primary outcome is turbulence, then planetesimal formation through the classic SI may be less efficient than previously thought, as clumping will always be hindered by small-scale VSSI-turbulence. High-resolution simulations that fully resolve VSSI scales will be necessary to clarify this issue.}


\acknowledgements
{I thank the anonymous referee for a thorough report that considerably improved the connection between this work and the  literature.} I thank Volker Elling, Pin-Gao Gu, Ming-Chih Lai, Yueh-Ning Lee, Te-Sheng Lin, Jack Ng, Debanjan Sengupta, {Ryosuke Tominaga,} Orkan Umurhan, and David C.C. Yen, for useful discussions, tips, and advice. This work is supported by Taiwan's Ministry of Science and Education through grant 107-2112-M001-043-MY3.

\appendix
{
\section{List of symbols}\label{appendix_symbols}

Table \ref{symbols} summarizes frequently used and related symbols in the main text.

\begin{deluxetable}{cll}
\tablecaption{{Frequently used symbols} \label{symbols}}
\tablehead{\colhead{Notation} & \colhead{Definition} & \colhead{Description}}
\startdata
$\rho_\mathrm{d,g}$ & & Dust and gas densities\\
$\bm{v}_\mathrm{d,g}$ & & Dust and gas velocities in the shearing box, relative to Keplerian flow\\
$\epsilon, \epsilon_0$ & $\rhod/\rhog$, $\epsilon(z=0)$ & Local dust-to-gas ratio, midplane dust-to-gas ratio \\
$\Sigma_\mathrm{d,g}$ & $\int_{-\infty}^\infty\rho_\mathrm{d,g}dz$ & Dust and gas surface densities\\
$Z$ & $\sigd/\sigg$ & Metallicity  \\
$c_s$ & & Constant gas sound-speed\\
$\Omega$ & $\sqrt{GM_*/r^3}$ & Keplerian rotation frequency\\
$\Hgas$, $\hgas$&  $c_s/\Omega$, $\Hgas/r$& Gas disk pressure scale-height, aspect-ratio \\
$P$ & $c_s^2\rhog$ & Pressure in the global disk or local pressure fluctuations in the shearing box \\
$\eta$, $\hat{\eta}$ & $-\left(r\Omega^2\rhog\right)^{-1}\p_r P$, $\eta/\hgas$ & Dimensionless global pressure gradient, reduced pressure gradient parameter \\
$\alpha$ & & Dimensionless gas viscosity\\
$\delta$ & $\left.\left(1 + \st + 4\st^2\right)\right/\left(1 + \st^2\right)^2\alpha$ & Dimensionless dust diffusion coefficient \\
$\st$ & $\taus\Omega$& Stokes number with particle stopping time $\taus$\\
$\Hdust$ & $\sqrt{\delta/\left(\delta + \st\right)}\Hgas$ & Dust scale-height\\
$\delta\rhog$, etc. & & Complex amplitude of Eulerian perturbations (eigenfunctions) \\
$s$, $\omega$ & $\re(\sigma)$, $-\im(\sigma)$ & Growth rate, oscillation frequency of the complex growth rate $\sigma$\\
$K_x$ & $k_x\Hgas$& Dimensionless radial wavenumber
\enddata
\end{deluxetable}
}

\section{One-fluid model of dusty gas in the shearing box}\label{one_fluid_model}


In the `one-fluid' description of dusty-gas we work with the total density
\begin{align}
    \rho = \rhog + \rhod
\end{align}
and center-of-mass velocity
\begin{align}
    \bm{v}_c = \frac{\rhog\vg + \rhod\vd}{\rho}.\label{com_1fluid}
\end{align}
Furthermore, by considering small, tightly-coupled dust particles we relate the gas and dust velocities by the `terminal velocity approximation',
\begin{align}
    \vd = \vg + \tstop \left( \frac{\nabla P}{\rhog} - 2\eta r\Omega^2 \hat{\bm{x}}\right)\label{terminal_velocity}
\end{align}
\citep{youdin05a,laibe14}. Here, $\tstop = \taus\rhog/\rho$ is the \emph{relative} stopping time. Recall that $\bm{v}_\mathrm{d,g}$ are dust and gas velocities in the shearing box relative to the Keplerian flow, respectively, and $P$ is the local pressure fluctuation. The term $\propto\eta$ represents the radial pressure gradient in the global disk. Thus, in the unperturbed state with vanishing $P$, dust drifts radially relative to the gas.

The dust-gas mixture is modeled as a single, adiabatic fluid with a special cooling function \citep{lin17,lovascio19}. Dropping the subscript `c' for clarity, our one-fluid model equations in the local shearing box
to $O(\tstop)$ are
\begin{align}
    &\frac{\p\rho}{\p t} + \nabla\cdot\left(\rho\bmv\right)
    = \nabla\cdot\left(D\rhog\nabla\epsilon\right),\label{1f_mass}\\
    &\frac{\p\bmv}{\p t} + \bmv\cdot\nabla\bmv = -\frac{\nabla P}{\rho} + 2\eta r\Omega^2 \frac{\rho_\mathrm{g}}{\rho}\hat{\bm{x}}  + 2\Omega v_y \hat{\bm{x}}
    -\frac{\Omega}{2} v_x \hat{\bm{y}} - \Omega^2 z\hat{\bm{z}},\label{1f_mom}\\
    &\frac{\p P}{\p t} + \nabla\cdot\left(P\bmv\right) = c_s^2\nabla\cdot\left[\tstop\fdust\left(\nabla P - 2 \eta r \Omega^2\rho_\mathrm{g}\hat{\bm{x}}\right)\right].
    \label{1f_energy}
\end{align}
{\citep{laibe14,lin17,lovascio19,chen20,paardekooper20}}, where $\fdust \equiv \rhod/\rho$ is the dust fraction. Note that $\rhog = P/c_s^2$ for the isothermal gas we consider, so $\fdust = 1 - P/c_s^2\rho$. The second term on the RHS of Eq. \ref{1f_mom} vanishes in a particle disk where $\rho_\mathrm{g}\to 0$, since solids do not feel pressure gradients. Conversely, for a dust-free gas disk ($\rho_\mathrm{g}/\rho \to 1$) we recover the full pressure support from the global disk. The second term in the parenthesis on the RHS of Eq. \ref{1f_energy} arises from the contribution to the terminal velocity approximation from the large-scale radial pressure gradient in Eq. \ref{terminal_velocity}.

\subsection{Approximate equilibria}
As in the two-fluid model we seek steady, horizontally uniform solutions. The mass, energy, and vertical momentum equations are then
\begin{align}
    \rho v_z &= D \rhod\dln{\epsilon},\\
    v_zv_z^\prime &= - \frac{P^\prime}{\rho} - \Omega^2 z \simeq 0,\label{1f_vz_eqm}\\
    Pv_z &= c_s^2 \tstop\fdust P^\prime,
\end{align}
where in Eq. \ref{1f_vz_eqm} we neglect the $O(v_z^2)$ term \emph{a posterior} for consistency with the small $\tstop$ approximation used to derive  the one-fluid model. These equations may then be solved for {constant $\tau_s = \tstop/\fg = \st/\Omega$} to yield
\begin{align}
    \epsilon & = \epsilon_0 \exp{\left(-\frac{\st}{2\delta}\frac{z^2}{\Hgas^2}\right)}, \label{1fluid_eqm_eps}\\
    v_z &= -\frac{\epsilon}{1 + \epsilon} \st z\Omega,\label{1fluid_eqm_vz}\\
    P &= P_0\exp{\left[\frac{\delta}{\st}\left(\epsilon - \epsilon_0\right) - \frac{z^2}{2\Hgas^2}\right]}.\label{1fluid_eqm_energy}
\end{align}
From here it is clear that $v_z = O(\st)$, so it is self-consistent to neglect the $O(v_z^2)$ term. 
Eq. \ref{1fluid_eqm_eps}--\ref{1fluid_eqm_energy}
are in fact the same solutions as in the full two fluid model in the limit $\st \ll 1$ (see Eq. \ref{2f_eps}--\ref{2f_vdz}, recall $\beta \to \st$ for $\st\to 0$; and note $v_z = \fdust\vdz $).

The horizontal momentum equations are
\begin{align}
    &v_z v_x^\prime  = 2 \eta r \Omega^2 \frac{\rhog}{\rho} + 2\Omega v_y \simeq 0,\label{1fluid_eqm_vx}\\
    &v_z v_y^\prime = - \frac{\Omega}{2}v_x, \label{1fluid_eqm_vy}
\end{align}
where we neglect the quadratic term in Eq. \ref{1fluid_eqm_vx} to obtain
\begin{align}
    v_y &= -\frac{\eta r \Omega}{1 + \epsilon}. \label{1fluid_eqm_vy_sol}
\end{align}
This is expected on physical grounds for tightly-coupled dust ($\st\to 0$). In this limit the mixture behaves close to a single fluid with orbital velocity depending on the level of dust enrichment. For $\epsilon \to 0$ we have a pressure-supported gas disk at sub-Keplerian velocity (assuming $\eta>0$); while $\epsilon\to\infty$ corresponds to a particle disk on exactly Keplerian orbits, since then $v_y\to 0$.

Next, we use Eq. \ref{1fluid_eqm_vy_sol}, Eq. \ref{1fluid_eqm_vy}, and Eq. \ref{1fluid_eqm_vz} to obtain
\begin{align}
    v_x = -\frac{2}{\Omega}v_z v_y^\prime = \frac{2 \eta r \Omega\epsilon\epsilon^\prime}{(1+\epsilon)^3}\st z.  \label{vx_1f}
\end{align}
Notice the radial velocity of the dust-gas mixture's center-of-mass depends on height. It is only zero at the midplane and for $|z|\to\infty$ where $\epsilon \to 0$. For $\eta>0$ the specific angular momentum decreases with increasing $|z|$: the mixture gains pressure support as it becomes more gas-rich away from the midplane. This means that as a parcel of the mixture settles, it finds itself having an angular momentum deficit compared to its surrounding; it thus drifts inwards ($v_x<0$), as indicated by Eq. \ref{vx_1f}.





\subsection{Linearized equations}
We linearize the one-fluid equations about the above basic state, with non-uniform $v_x(z),\, v_y(z),$ and $v_z(z)$. As in the main text we assume axisymmetric perturbations in the form of $\delta\rho(z)\exp{\left(\sigma t + \ii k_x x\right)}$, and similarly for other variables. The linearized equations are

\begin{align}
    &\sigma\delrho  + \ii k_x \left(\dd v_x + v_x \delrho\right) + \frac{\rho^\prime}{\rho}\left(v_z\delrho + \dd v_z\right) + v_z\left(\delrho\right)^\prime + v_z^\prime \delrho + \dd v_z^\prime \notag \\
    &= D\fdust\left[
    Q^{\prime\prime}-k_x^2Q + \frac{\rhod^\prime}{\rhod}
    \left(\frac{\epsilon^\prime}{\epsilon}\delrhod + Q^\prime\right) + \frac{\epsilon^\prime}{\epsilon}\left(\delrhod\right)^\prime + (\ln\epsilon)^{\prime\prime}\delrhod
    \right],\\
    &\sigma\dd v_x + \ii k_x v_x \dd v_x + v_x^\prime\dd v_z  + v_z \dd v_x^\prime = -\ii k_x\frac{P}{\rho}W
    - \frac{2\eta r \Omega^2\epsilon}{(1+\epsilon)^2}Q
    + 2\Omega\dd v_y,\label{1fluid_lin_vx}\\
    &\sigma\dd v_y + \ii k_x v_x \dd v_y + v_y^\prime\dd v_z  + v_z\dd v_y^\prime = - \frac{\Omega}{2}\dd v_x,\label{1fluid_lin_vy}\\
    & \sigma\dd v_z + \ii k_x v_x \dd v_z + v_z^\prime\dd v_z + v_z \dd v_z^\prime = \frac{P^\prime}{\rho}\frac{\epsilon Q}{1+\epsilon}  - \frac{P}{\rho}W^\prime,\label{1fluid_lin_vz}\\
    & \sigma W  + \ii k_x\left(\dd v_x + v_x W\right) + \frac{P^\prime}{P}\left(v_z W + \dd v_z\right)+ v_z^\prime W + v_zW^\prime + \dd v_z^\prime\notag\\
    &= \frac{\calK}{P}\left[
    W^{\prime\prime} - k_x^2W + \frac{\calK^\prime}{\calK}\left(\frac{P^\prime}{P}\delK + W^\prime\right) + \frac{P^\prime}{P} \left(\delK\right)^\prime + \left(\ln P\right)^{\prime\prime}\delK
    \right] \notag\\
    &\phantom{=} - 2\ii k_x  \eta r \Omega^2\frac{\calK}{c_s^2 P} \left[\left(\frac{1-\epsilon}{1+\epsilon}\right)Q + W\right], \label{one_fluid_energy_linear}
\end{align}
where
\begin{align}
    \mathcal{K} &\equiv \frac{c_s^2P\st\epsilon}{\Omega(1+\epsilon)^2},
\end{align}
and recall $Q = \dd\epsilon/\epsilon$ and $W = \dd\rhog/\rhog = \dd P/P$.

\subsection{Mode energetics}\label{pseudo_energy}
Following \cite{ishitsu09}, we multiply the $x$, $y$, and $z$ momentum equations (Eq. \ref{1fluid_lin_vx}--\ref{1fluid_lin_vz}) by $\dd v_{x,y,z}^*$, respectively, combine them appropriately, then take the real part. We also scale the overall result by a factor of $(1+\epsilon)$ for easier comparison with the corresponding two-fluid treatment in Appendix \ref{2fluid_energy}. The one-fluid result is


\begin{align}
  E_\mathrm{tot} \equiv
  (1+\epsilon)\left(\left|\dd v_x\right|^2 + 4\left|\dd v_y\right|^2 + \left|\dd v_z\right|^2\right)
  = \sum_{i=1}^{5}E_i,
\end{align}
with
\begin{align}
    sE_1 &= -(1+\epsilon)\left[v_x^\prime \real{\dd v_z \dd v_x^*}
    + 4v_y^\prime \real{\delta v_z\dd v_y^*} + v_z^\prime\left|\dd v_z\right|^2
    \right]\\
    &\equiv sE_{1x} + sE_{1y} + sE_{1z}\notag\\
    sE_2 &= -v_z(1+\epsilon)\real{\delta v_x^\prime \dd v_x^* + 4\delta v_y^\prime \dd v_y^* + \delta v_z^\prime \dd v_z^*}\\
    sE_3 &= c_s^2\left[k_x \imag{W\dd v_x^*} - \real{W^\prime\dd v_z^*}\right]\\
    sE_4 &= - \frac{2\eta r\Omega^2\epsilon}{(1+\epsilon)}\real{Q\dd v_x^*} = \frac{\epsilon\Omega\left(\vdx - \vgx\right)}{\st}\real{Q\dd v_x^*}\\
    sE_5 &= -\epsilon z\Omega^2\real{Q\dd v_z^*} = - z\Omega^2(1+\epsilon)\operatorname{Re}\left[\left(\frac{\dd\rho}{\rho} - \frac{\delta P}{P}\right)\dd v_z^*\right].
\end{align}

The factor of 4 in the expression for $E_\mathrm{tot}$ is introduced to eliminate the rotation terms in Eq. \ref{1fluid_lin_vx} and \ref{1fluid_lin_vy}. $E_1$ is associated with the vertical shear in the equilibrium velocities, $E_2$ is associated with dust settling, and $E_3$ is associated with pressure forces. In the second equality for $E_4$, we used Eq. \ref{terminal_velocity} to relate the global radial pressure gradient to the dust-gas radial drift in steady state. For $E_5$, we used the equilibrium condition $P^\prime/\rho \simeq -z\Omega^2$; and in the second equality used fact that the total density $\rho\propto P(1+\epsilon)$ to relate the perturbed dust-to-gas ratio to pressure and density perturbations. From this it is clear that $E_5$ is associated with buoyancy effects, i.e. pressure-density perturbation mismatches.


\section{Two-fluid pseudo-energy decomposition}\label{2fluid_energy}

Following the same procedure as in the one-fluid treatment (Appendix \S\ref{pseudo_energy}), we can define the pseudo-energy in the full two-fluid framework as
\begin{align}
    U_\mathrm{tot} = \epsilon\left(
    \left|\dd \vdx \right|^2 +  4\left|\dd \vdy \right|^2 + \left|\dd \vdz \right|^2\right) +
    \left|\dd \vgx \right|^2 +  4\left|\dd \vgy \right|^2 + \left|\dd \vgz \right|^2
     = \sum_{i=1}^6 U_i,
\end{align}
with
\begin{align}
& s U_1 = -\left[\epsilon\vdx^\prime\real{\dd\vdz\dd\vdx^*} + \vgx^\prime\real{\dd\vgz\dd\vgx^*}\right] - 4\left[\epsilon \vdy^\prime\real{\dd\vdz\dd\vdy^*} + \vgy^\prime\real{\dd\vgz\dd\vgy^*}\right] - \epsilon\vdz^\prime\left|\dd\vdz\right|^2,\\
&\phantom{sU_1} \equiv sU_{1x} + s U_{1y} + s U_{1z}, \notag \\
& sU_2 = - \epsilon \vdz \real{\dd\vdx^\prime\dd\vdx^* + 4\dd\vdy^\prime\dd\vdy^* + \dd\vdz^\prime\dd\vdz^*},\\
&sU_3 = k_x c_s^2 \imag{W\dd\vgx^*} - c_s^2 \real{W^\prime \delta\vgz^*},\\
&s U_4 =-\frac{\epsilon\Omega}{\st}\left[ \left(\vgx - \vdx\right)\real{Q\dd\vgx^*} + 4 \left(\vgy - \vdy\right)\real{Q\dd\vgy^*}\right.\notag\\
&\phantom{sU_4=-\frac{\epsilon\Omega}{\st}\left[\right.}
+\left. \left|\dd\vgx - \dd\vdx\right|^2 + 4\left|\dd\vgx - \dd\vdx\right|^2 + \left|\dd\vgz - \dd\vdz\right|^2
\right],\\
& sU_5 = \frac{\epsilon\Omega}{\st}\vdz \real{Q\dd\vgz^*},\\
&sU_6 = \real{\dd F^\mathrm{visc}_x \dd\vgx^* + 4\dd F^\mathrm{visc}_y \dd\vgy^* + \dd F^\mathrm{visc}_z \dd\vgz^*},
\end{align}
where $\delta\bm{F}^\mathrm{visc}$ is given by Eq. \ref{dFvisc_x}--\ref{dFvisc_z}.

As in the one-fluid model, we can associate $U_1$ with the vertical shear in the equilibrium velocities; $U_2$ with dust settling, $U_3$ with gas pressure forces, $U_4$ with the dust-gas relative drift, and $U_5$ with vertical buoyancy.
The full two-fluid framework also includes viscous contributions,  $U_6$, which is neglected in the one-fluid treatment.

\bibliographystyle{aasjournal}
\bibliography{ref}

\begin{thebibliography}{}
\expandafter\ifx\csname natexlab\endcsname\relax\def\natexlab#1{#1}\fi
\providecommand{\url}[1]{\href{#1}{#1}}

\bibitem[{{Adams} {et~al.}(1989){Adams}, {Ruden}, \& {Shu}}]{adams89}
{Adams}, F.~C., {Ruden}, S.~P., \& {Shu}, F.~H. 1989, \apj, 347, 959

\bibitem[{{Auffinger} \& {Laibe}(2018)}]{auffinger18}
{Auffinger}, J., \& {Laibe}, G. 2018, \mnras, 473, 796

\bibitem[{{Bai} \& {Stone}(2010{\natexlab{a}})}]{bai10b}
{Bai}, X.-N., \& {Stone}, J.~M. 2010{\natexlab{a}}, \apjs, 190, 297

\bibitem[{{Bai} \& {Stone}(2010{\natexlab{b}})}]{bai10}
---. 2010{\natexlab{b}}, \apj, 722, 1437

\bibitem[{{Bai} \& {Stone}(2010{\natexlab{c}})}]{bai10c}
---. 2010{\natexlab{c}}, \apjl, 722, L220

\bibitem[{{Balbus}(2003)}]{balbus03}
{Balbus}, S.~A. 2003, \araa, 41, 555

\bibitem[{{Balbus} \& {Hawley}(1992)}]{balbus92}
{Balbus}, S.~A., \& {Hawley}, J.~F. 1992, \apj, 400, 610

\bibitem[{Balsara {et~al.}(2009)Balsara, Tilley, Rettig, \&
  Brittain}]{balsara09}
Balsara, D.~S., Tilley, D.~A., Rettig, T., \& Brittain, S.~D. 2009, Monthly
  Notices of the Royal Astronomical Society, 397, 24.
\newblock \url{https://doi.org/10.1111/j.1365-2966.2009.14606.x}

\bibitem[{{Barker} \& {Latter}(2015)}]{barker15}
{Barker}, A.~J., \& {Latter}, H.~N. 2015, \mnras, 450, 21

\bibitem[{{Ben{\'\i}tez-Llambay} {et~al.}(2019){Ben{\'\i}tez-Llambay}, {Krapp},
  \& {Pessah}}]{llambay19}
{Ben{\'\i}tez-Llambay}, P., {Krapp}, L., \& {Pessah}, M.~E. 2019, \apjs, 241,
  25

\bibitem[{{Birnstiel} {et~al.}(2016){Birnstiel}, {Fang}, \&
  {Johansen}}]{birnstiel16}
{Birnstiel}, T., {Fang}, M., \& {Johansen}, A. 2016, \ssr, 205, 41

\bibitem[{{Birnstiel} {et~al.}(2012){Birnstiel}, {Klahr}, \&
  {Ercolano}}]{birnstiel12}
{Birnstiel}, T., {Klahr}, H., \& {Ercolano}, B. 2012, \aap, 539, A148

\bibitem[{{Blum}(2018)}]{blum18}
{Blum}, J. 2018, \ssr, 214, 52

\bibitem[{{Burns} {et~al.}(2019){Burns}, {Vasil}, {Oishi}, {Lecoanet}, \&
  {Brown}}]{burns19}
{Burns}, K.~J., {Vasil}, G.~M., {Oishi}, J.~S., {Lecoanet}, D., \& {Brown},
  B.~P. 2019, arXiv e-prints, arXiv:1905.10388

\bibitem[{{Carrera} {et~al.}(2020){Carrera}, {Simon}, {Li}, {Kretke}, \&
  {Klahr}}]{carrera20}
{Carrera}, D., {Simon}, J.~B., {Li}, R., {Kretke}, K.~A., \& {Klahr}, H. 2020,
  arXiv e-prints, arXiv:2008.01727

\bibitem[{{Chen} \& {Lin}(2020)}]{chen20}
{Chen}, K., \& {Lin}, M.-K. 2020, \apj, 891, 132

\bibitem[{{Chiang}(2008)}]{chiang08}
{Chiang}, E. 2008, \apj, 675, 1549

\bibitem[{{Chiang} \& {Youdin}(2010)}]{chiang10}
{Chiang}, E., \& {Youdin}, A.~N. 2010, Annual Review of Earth and Planetary
  Sciences, 38, 493

\bibitem[{{Dubrulle} {et~al.}(1995){Dubrulle}, {Morfill}, \&
  {Sterzik}}]{dubruelle95}
{Dubrulle}, B., {Morfill}, G., \& {Sterzik}, M. 1995, \icarus, 114, 237

\bibitem[{{Flock} {et~al.}(2017){Flock}, {Nelson}, {Turner}, {Bertrang},
  {Carrasco-Gonz{\'a}lez}, {Henning}, {Lyra}, \& {Teague}}]{flock17}
{Flock}, M., {Nelson}, R.~P., {Turner}, N.~J., {et~al.} 2017, \apj, 850, 131

\bibitem[{{Garaud} \& {Lin}(2004)}]{garaud04}
{Garaud}, P., \& {Lin}, D.~N.~C. 2004, \apj, 608, 1050

\bibitem[{{Goldreich} \& {Lynden-Bell}(1965)}]{goldreich65}
{Goldreich}, P., \& {Lynden-Bell}, D. 1965, \mnras, 130, 125

\bibitem[{{Goldreich} \& {Ward}(1973)}]{goldreich73}
{Goldreich}, P., \& {Ward}, W.~R. 1973, \apj, 183, 1051

\bibitem[{{Gole} {et~al.}(2020){Gole}, {Simon}, {Li}, {Youdin}, \&
  {Armitage}}]{gole20}
{Gole}, D.~A., {Simon}, J.~B., {Li}, R., {Youdin}, A.~N., \& {Armitage}, P.~J.
  2020, arXiv e-prints, arXiv:2001.10000

\bibitem[{{Ishitsu} {et~al.}(2009){Ishitsu}, {Inutsuka}, \&
  {Sekiya}}]{ishitsu09}
{Ishitsu}, N., {Inutsuka}, S.-i., \& {Sekiya}, M. 2009, arXiv e-prints,
  arXiv:0905.4404

\bibitem[{{Jacquet} {et~al.}(2011){Jacquet}, {Balbus}, \& {Latter}}]{jacquet11}
{Jacquet}, E., {Balbus}, S., \& {Latter}, H. 2011, \mnras, 415, 3591

\bibitem[{{Jaupart} \& {Laibe}(2020)}]{jaupart20}
{Jaupart}, E., \& {Laibe}, G. 2020, \mnras, 492, 4591

\bibitem[{{Johansen} {et~al.}(2014){Johansen}, {Blum}, {Tanaka}, {Ormel},
  {Bizzarro}, \& {Rickman}}]{johansen14}
{Johansen}, A., {Blum}, J., {Tanaka}, H., {et~al.} 2014, Protostars and Planets
  VI, 547

\bibitem[{{Johansen} \& {Youdin}(2007)}]{johansen07}
{Johansen}, A., \& {Youdin}, A. 2007, \apj, 662, 627

\bibitem[{{Johansen} {et~al.}(2009){Johansen}, {Youdin}, \& {Mac
  Low}}]{johansen09}
{Johansen}, A., {Youdin}, A., \& {Mac Low}, M.-M. 2009, \apjl, 704, L75

\bibitem[{{Johnson} \& {Gammie}(2005)}]{johnson05}
{Johnson}, B.~M., \& {Gammie}, C.~F. 2005, \apj, 626, 978

\bibitem[{{Kowalik} {et~al.}(2013){Kowalik}, {Hanasz}, {W{\'o}lta{\'n}ski}, \&
  {Gawryszczak}}]{kowalik13}
{Kowalik}, K., {Hanasz}, M., {W{\'o}lta{\'n}ski}, D., \& {Gawryszczak}, A.
  2013, \mnras, 434, 1460

\bibitem[{{Krapp} {et~al.}(2019){Krapp}, {Ben{\'\i}tez-Llambay}, {Gressel}, \&
  {Pessah}}]{krapp19}
{Krapp}, L., {Ben{\'\i}tez-Llambay}, P., {Gressel}, O., \& {Pessah}, M.~E.
  2019, \apjl, 878, L30

\bibitem[{{Krapp} {et~al.}(2020){Krapp}, {Youdin}, {Kratter}, \&
  {Ben{\'\i}tez-Llambay}}]{krapp20}
{Krapp}, L., {Youdin}, A.~N., {Kratter}, K.~M., \& {Ben{\'\i}tez-Llambay}, P.
  2020, \mnras, arXiv:2004.04590

\bibitem[{{Laibe} {et~al.}(2020){Laibe}, {Br{\'e}hier}, \& {Lombart}}]{laibe20}
{Laibe}, G., {Br{\'e}hier}, C.-E., \& {Lombart}, M. 2020, \mnras, 494, 5134

\bibitem[{{Laibe} \& {Price}(2014)}]{laibe14}
{Laibe}, G., \& {Price}, D.~J. 2014, \mnras, 440, 2136

\bibitem[{{Latter} \& {Ogilvie}(2006)}]{latter06}
{Latter}, H.~N., \& {Ogilvie}, G.~I. 2006, \mnras, 372, 1829

\bibitem[{{Latter} \& {Papaloizou}(2018)}]{latter18}
{Latter}, H.~N., \& {Papaloizou}, J. 2018, \mnras, 474, 3110

\bibitem[{{Lee} {et~al.}(2010){Lee}, {Chiang}, {Asay-Davis}, \&
  {Barranco}}]{lee10}
{Lee}, A.~T., {Chiang}, E., {Asay-Davis}, X., \& {Barranco}, J. 2010, \apj,
  718, 1367

\bibitem[{{Li} {et~al.}(2019){Li}, {Youdin}, \& {Simon}}]{li19}
{Li}, R., {Youdin}, A.~N., \& {Simon}, J.~B. 2019, \apj, 885, 69

\bibitem[{{Lin}(2013)}]{lin13}
{Lin}, M.-K. 2013, \apj, 765, 84

\bibitem[{{Lin}(2019)}]{lin19}
---. 2019, \mnras, 485, 5221

\bibitem[{{Lin} \& {Kratter}(2016)}]{lin16}
{Lin}, M.-K., \& {Kratter}, K.~M. 2016, \apj, 824, 91

\bibitem[{{Lin} \& {Papaloizou}(2011{\natexlab{a}})}]{lin11a}
{Lin}, M.-K., \& {Papaloizou}, J.~C.~B. 2011{\natexlab{a}}, \mnras, 415, 1426

\bibitem[{{Lin} \& {Papaloizou}(2011{\natexlab{b}})}]{lin11b}
---. 2011{\natexlab{b}}, \mnras, 415, 1445

\bibitem[{{Lin} \& {Youdin}(2015)}]{lin15}
{Lin}, M.-K., \& {Youdin}, A.~N. 2015, \apj, 811, 17

\bibitem[{{Lin} \& {Youdin}(2017)}]{lin17}
---. 2017, \apj, 849, 129

\bibitem[{{Lovascio} \& {Paardekooper}(2019)}]{lovascio19}
{Lovascio}, F., \& {Paardekooper}, S.-J. 2019, \mnras, 488, 5290

\bibitem[{{Lubow} \& {Pringle}(1993)}]{lubow93}
{Lubow}, S.~H., \& {Pringle}, J.~E. 1993, \apj, 409, 360

\bibitem[{{Manger} {et~al.}(2020){Manger}, {Klahr}, {Kley}, \&
  {Flock}}]{manger20}
{Manger}, N., {Klahr}, H., {Kley}, W., \& {Flock}, M. 2020, \mnras, 499, 1841

\bibitem[{{Mathis} {et~al.}(1977){Mathis}, {Rumpl}, \& {Nordsieck}}]{mathis77}
{Mathis}, J.~S., {Rumpl}, W., \& {Nordsieck}, K.~H. 1977, \apj, 217, 425

\bibitem[{{McNally} \& {Pessah}(2014)}]{mcnally14}
{McNally}, C.~P., \& {Pessah}, M.~E. 2014, ArXiv e-prints, arXiv:1406.4864

\bibitem[{{Nakagawa} {et~al.}(1986){Nakagawa}, {Sekiya}, \&
  {Hayashi}}]{nakagawa86}
{Nakagawa}, Y., {Sekiya}, M., \& {Hayashi}, C. 1986, \icarus, 67, 375

\bibitem[{{Nelson} {et~al.}(2013){Nelson}, {Gressel}, \& {Umurhan}}]{nelson13}
{Nelson}, R.~P., {Gressel}, O., \& {Umurhan}, O.~M. 2013, \mnras, 435, 2610

\bibitem[{{Nesvorn{\'y}} {et~al.}(2019){Nesvorn{\'y}}, {Li}, {Youdin}, {Simon},
  \& {Grundy}}]{nesvorny19}
{Nesvorn{\'y}}, D., {Li}, R., {Youdin}, A.~N., {Simon}, J.~B., \& {Grundy},
  W.~M. 2019, Nature Astronomy, 364

\bibitem[{{Paardekooper} {et~al.}(2020){Paardekooper}, {McNally}, \&
  {Lovascio}}]{paardekooper20}
{Paardekooper}, S.-J., {McNally}, C.~P., \& {Lovascio}, F. 2020, arXiv
  e-prints, arXiv:2010.01145

\bibitem[{{Pan}(2020)}]{pan20b}
{Pan}, L. 2020, \apj, 898, 8

\bibitem[{{Pan} \& {Yu}(2020)}]{pan20a}
{Pan}, L., \& {Yu}, C. 2020, \apj, 898, 7

\bibitem[{{Price} \& {Laibe}(2015)}]{price15}
{Price}, D.~J., \& {Laibe}, G. 2015, \mnras, 451, 813

\bibitem[{{Savonije} \& {Heemskerk}(1990)}]{savonije90}
{Savonije}, G.~J., \& {Heemskerk}, M.~H.~M. 1990, \aap, 240, 191

\bibitem[{{Sch{\"a}fer} {et~al.}(2020){Sch{\"a}fer}, {Johansen}, \&
  {Banerjee}}]{schafer20}
{Sch{\"a}fer}, U., {Johansen}, A., \& {Banerjee}, R. 2020, \aap, 635, A190

\bibitem[{{Sch{\"a}fer} {et~al.}(2017){Sch{\"a}fer}, {Yang}, \&
  {Johansen}}]{schafer17}
{Sch{\"a}fer}, U., {Yang}, C.-C., \& {Johansen}, A. 2017, \aap, 597, A69

\bibitem[{{Schaffer} {et~al.}(2018){Schaffer}, {Yang}, \&
  {Johansen}}]{schaffer18}
{Schaffer}, N., {Yang}, C.-C., \& {Johansen}, A. 2018, \aap, 618, A75

\bibitem[{{Schreiber} \& {Klahr}(2018)}]{schreiber18}
{Schreiber}, A., \& {Klahr}, H. 2018, \apj, 861, 47

\bibitem[{{Shakura} \& {Sunyaev}(1973)}]{shakura73}
{Shakura}, N.~I., \& {Sunyaev}, R.~A. 1973, \aap, 24, 337

\bibitem[{{Shi} \& {Chiang}(2013)}]{shi13}
{Shi}, J.-M., \& {Chiang}, E. 2013, \apj, 764, 20

\bibitem[{{Simon} {et~al.}(2016){Simon}, {Armitage}, {Li}, \&
  {Youdin}}]{simon16}
{Simon}, J.~B., {Armitage}, P.~J., {Li}, R., \& {Youdin}, A.~N. 2016, \apj,
  822, 55

\bibitem[{{Squire} \& {Hopkins}(2018{\natexlab{a}})}]{squire18a}
{Squire}, J., \& {Hopkins}, P.~F. 2018{\natexlab{a}}, \apjl, 856, L15

\bibitem[{{Squire} \& {Hopkins}(2018{\natexlab{b}})}]{squire18b}
---. 2018{\natexlab{b}}, \mnras, 477, 5011

\bibitem[{{Squire} \& {Hopkins}(2020)}]{squire20}
---. 2020, \mnras, 498, 1239

\bibitem[{{Stoll} \& {Kley}(2016)}]{stoll16}
{Stoll}, M. H.~R., \& {Kley}, W. 2016, \aap, 594, A57

\bibitem[{{Takeuchi} \& {Lin}(2002)}]{takeuchi02}
{Takeuchi}, T., \& {Lin}, D.~N.~C. 2002, \apj, 581, 1344

\bibitem[{{Testi} {et~al.}(2014){Testi}, {Birnstiel}, {Ricci}, {Andrews},
  {Blum}, {Carpenter}, {Dominik}, {Isella}, {Natta}, {Williams}, \&
  {Wilner}}]{testi14}
{Testi}, L., {Birnstiel}, T., {Ricci}, L., {et~al.} 2014, in Protostars and
  Planets VI, ed. H.~{Beuther}, R.~S. {Klessen}, C.~P. {Dullemond}, \&
  T.~{Henning}, 339

\bibitem[{{Tilley} {et~al.}(2010){Tilley}, {Balsara}, {Brittain}, \&
  {Rettig}}]{tilley11}
{Tilley}, D.~A., {Balsara}, D.~S., {Brittain}, S.~D., \& {Rettig}, T. 2010,
  \mnras, 403, 211

\bibitem[{{Tominaga} {et~al.}(2019){Tominaga}, {Takahashi}, \&
  {Inutsuka}}]{tominaga19}
{Tominaga}, R.~T., {Takahashi}, S.~Z., \& {Inutsuka}, S.-i. 2019, \apj, 881, 53

\bibitem[{{Umurhan} {et~al.}(2020){Umurhan}, {Estrada}, \& {Cuzzi}}]{umurhan20}
{Umurhan}, O.~M., {Estrada}, P.~R., \& {Cuzzi}, J.~N. 2020, \apj, 895, 4

\bibitem[{{Weidenschilling}(1977)}]{weiden77}
{Weidenschilling}, S.~J. 1977, \mnras, 180, 57

\bibitem[{{Whipple}(1972)}]{whipple72}
{Whipple}, F.~L. 1972, in From Plasma to Planet, ed. A.~{Elvius}, 211

\bibitem[{{Yang} \& {Johansen}(2014)}]{yang14}
{Yang}, C.-C., \& {Johansen}, A. 2014, \apj, 792, 86

\bibitem[{{Yang} {et~al.}(2017){Yang}, {Johansen}, \& {Carrera}}]{yang17}
{Yang}, C.~C., {Johansen}, A., \& {Carrera}, D. 2017, \aap, 606, A80

\bibitem[{{Yang} {et~al.}(2018){Yang}, {Mac Low}, \& {Johansen}}]{yang18}
{Yang}, C.-C., {Mac Low}, M.-M., \& {Johansen}, A. 2018, \apj, 868, 27

\bibitem[{{Youdin} \& {Johansen}(2007)}]{youdin07b}
{Youdin}, A., \& {Johansen}, A. 2007, \apj, 662, 613

\bibitem[{{Youdin}(2011)}]{youdin11}
{Youdin}, A.~N. 2011, \apj, 731, 99

\bibitem[{{Youdin} \& {Goodman}(2005)}]{youdin05a}
{Youdin}, A.~N., \& {Goodman}, J. 2005, \apj, 620, 459

\bibitem[{{Youdin} \& {Lithwick}(2007)}]{youdin07}
{Youdin}, A.~N., \& {Lithwick}, Y. 2007, \icarus, 192, 588

\bibitem[{{Zhu} {et~al.}(2015){Zhu}, {Stone}, \& {Bai}}]{zhu15}
{Zhu}, Z., {Stone}, J.~M., \& {Bai}, X.-N. 2015, \apj, 801, 81

\bibitem[{{Zhu} \& {Yang}(2020)}]{zhu20}
{Zhu}, Z., \& {Yang}, C.-C. 2020, arXiv e-prints, arXiv:2008.01119

\bibitem[{{Zhuravlev}(2019)}]{zhuravlev19}
{Zhuravlev}, V.~V. 2019, \mnras, 489, 3850

\bibitem[{{Zhuravlev}(2020)}]{zhuravlev20}
---. 2020, \mnras, 494, 1395

\end{thebibliography}

\end{document}